\documentclass[fleqn,usenatbib]{mnras}

\usepackage[T1]{fontenc}
\usepackage{ae,aecompl}

\usepackage{graphicx}	
\usepackage{amsmath}	
\usepackage{amssymb}	
\usepackage{longtable}
\usepackage{appendix}
\usepackage{soul}
\usepackage{ulem}

\usepackage{cprotect}

\usepackage{newtxtext,newtxmath}

\usepackage[usenames, dvipsnames]{xcolor}
\usepackage{multirow}
\usepackage{lastpage}
\usepackage[export]{adjustbox}

\title[CAFOS DR1]{The Calar Alto CAFOS Direct Imaging First Data Release}

\author[Cort\'es-Contreras et al.]{
M. Cort\'es-Contreras,$^{1}$\thanks{E-mail: mcortes@cab.inta-csic.es}
E. Solano,$^{1}$
J. Alonso-Hern\'andez,$^{1,2}$
N. Cardiel,$^{2,3}$
P. Cruz,$^{1}$
\newauthor
C. Rodrigo$^{1}$
\\
$^{1}$Centro de Astrobiolog\'{\i}a (CAB), CSIC-INTA, Camino Bajo del Castillo s/n, campus ESAC, 28692, Villanueva de la Ca\~nada, Madrid, Spain\\
$^{2}$Departamento de F\'isica de la Tierra y Astrof\'isica, Fac. CC. F\'isicas, Universidad Complutense de Madrid, Plaza de las Ciencias 1, E-28040 Madrid, Spain\\
$^{3}$ Instituto de F\'isica de Part\'iculas y del Cosmos, IPARCOS, Fac. CC. F\'isicas, Universidad Complutense de MAdrid, Plaza de las Ciencias 1, E-28040 Madrid,Spain\\
}

\date{Accepted XXX. Received YYY; in original form ZZZ}

\pubyear{2021}

\begin{document}
\label{firstpage}
\pagerange{\pageref{firstpage}--\pageref{lastpage}}
\maketitle

\begin{abstract}
We present the first release of the Calar Alto CAFOS direct imaging data, a project led by the Spanish Virtual Observatory with the goal of enhancing the use of the Calar Alto archive by the astrophysics community. Data Release 1 contains 23\,903 reduced and astrometrically calibrated images taken from March 2008 to July 2019 with a median of the mean uncertainties in the astrometric calibration of 0.04\,arcsec. 
The catalogue associated to 6\,132 images in the Sloan $griz$ filters provides accurate astrometry and PSF calibrated photometry for 139\,337 point-like detections corresponding to 21\,985 different sources extracted from a selection of 2\,338 good-quality images. The mean internal astrometric and photometric accuracies are 0.05\,arcsec and 0.04\,mag, respectively In this work we describe the approach followed to process and calibrate the images, and the construction of the associated catalogue, together with the validation quality tests carried out. Finally, we present three cases to prove the science capabilities of the catalogue: discovery and identification of asteroids, identification of potential transients, and identification of cool and ultracool dwarfs.

\end{abstract}


\section{Introduction}

The large amount of data obtained from ground and space based telescopes make astronomical archives and catalogues prime assets for astrophysics research. They provide raw and reduced (photometrically and astrometrically corrected images, spectra ready for immediate scientific exploitation,...) data, but also high-level (catalogues, mosaics, stacked images,...) data products highly demanded by the scientific community.
Projects like SDSS \citep{York00}, 2MASS \citep{2003yCat.2246....0C}, UKIDSS \citep{UKIDSS2007_2} or WISE \citep{Wright10}, to name a few, are good examples of extensively used science-ready data products. Their archives are supported by a large number of refereed papers, which reflects their usefulness when conducting research projects. Archival data do also represent an advantage for the community in terms of time consumption, since they provide multiwavelength data that cover large regions of the sky without requiring new observations and their resulting processing time.

The Calar Alto Observatory (CAHA, Centro Astron\'{o}mico Hispano en Andaluc\'{\i}a) is the largest astronomical observatory located in continental Europe. Its 2.2\,m telescope hosts the Calar Alto Faint Object Spectrograph (CAFOS), a versatile instrument that allows four modes of observation: direct imaging, spectroscopy, polarimetry and narrow band imaging using a Fabry P\'erot etalon. In its standard configuration\textcolor{violet}{,} CAFOS is equipped with a 2048 × 2048 pixel blue sensitive CCD with an image scale of 0.53\,arcsec/pix.

The Virtual Observatory (VO) is a worldwide project that performs under the International Virtual Observatory Alliance\footnote{\url{www.ivoa.net}}. Its main goal is to develop a standardized framework that enables data centers to provide competing and co-operating data services, as well as powerful analysis, visualization tools and user interfaces to enhance the scientific exploitation of astronomical data. The Spanish Virtual Observatory (SVO\footnote{\url{https://svo.cab.inta-csic.es/main/index.php}}) is one of the 21 VO initiatives distributed worldwide. Among other data collections, SVO is responsible for the Calar Alto archive\footnote{\url{http://caha.sdc.cab.inta-csic.es/calto/index.jsp}}, in operation since 2011. 

This paper presents the first release of the CAFOS direct imaging data, which includes reduced and astrometrically calibrated images and its associated catalogue of point-like sources. We describe in Section~\ref{sec:data} the preparation and processing of the images. In Section~\ref{sec:astrometric_calibration}, we present the astrometric calibration followed by the source extraction and selection, and the photometric calibration in Sections~\ref{sec:sources} and \ref{sec:calibration}, respectively. The selection of images for building the catalogue is described in Section~\ref{sec:image_selection}. The point-like catalogue of sources and validation exercises carried out with Sloan Digital Sky Survey (SDSS) DR12 \citep{sdssdr12} data are introduced in Section~\ref{sec:pointlike}. We report two science cases carried out with the catalogue aiming at showing its potential for science exploitation in Section~\ref{sec:science}. We finish with a description on how to access the data and a summary in Sections~\ref{sec:data_access} and \ref{sec:conclusions}, respectively.

\section{Data}\label{sec:data}

\subsection{Data selection}\label{sec:data_preparation}

The sample of images to be processed is constituted by 42\,173 exposures observed with CAFOS between March 2008 and July 2019 that can be accessed from the Calar Alto archive.

For the first data release of the astrometric and photometric catalogue and to simplify, we will focus only on broad band images observed in the Sloan system of filters $g$ ($\lambda$ 4\,750\,\AA), $r$ ($\lambda$ 6\,300\,\AA), $i$ ($\lambda$ 7\,800\,\AA) and $z$ ($\lambda$ 9\,250\,\AA). There were a total of 6616 such images observed in the Sloan’s $griz$ bands between September 2012 and June 2019, with most of them ($\sim$ 75\%) taken between 2017 and 2018. Another 34607 images were observed with Johnson, Cousin or Gunn filters, among others, that will be detailed in a later data release. There were a further 950 images with no filter information in their headers.

\begin{figure*}
\centering
\includegraphics[width=\textwidth]{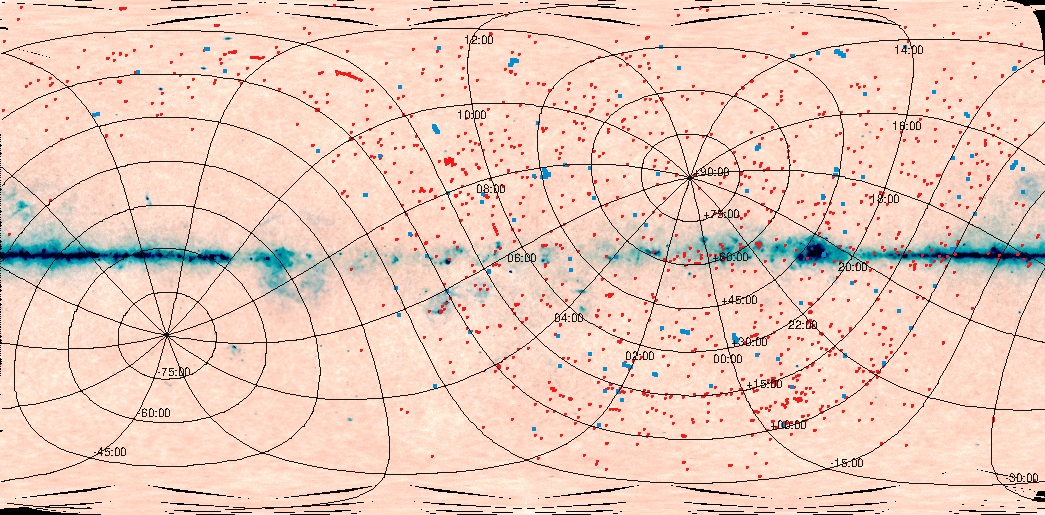}
\cprotect\caption{Location of the 42\,173 (red circles) and 6\,616 $griz$ (blue squares) CAFOS images in the sky. A Planck map is displayed in the background.}
\label{fig:skycoverage}
\end{figure*}

The location in the sky of the 42\,173 and 6\,616 images is shown in Figure~\ref{fig:skycoverage}. Due to the fact that CAFOS is not a survey instrument but, quite the contrary, devoted to diverse science projects, the sky distribution of the pointings is spread all across the northern sky and covers 22.8\,deg$^2$ of it (3.7\,deg$^2$ in the selection of images in the Sloan's filters).

\subsection{Data processing}\label{sec:data_processing}

Each of the 42\,173 images was processed using {\tt Filabres}\footnote{\url{https://filabres.readthedocs.io/en/latest/}}, a pipeline developed at the Universidad Complutense de Madrid jointly with the Calar Alto Observatory and the Spanish Virtual Observatory \citep{filabres_sea}. It is a software devoted to provide to the community useful images through the Calar Alto Archive. It is a three steps pipeline: (i) semi-automatic image classification into bias, flat-imaging, arc, science-imaging, etc., performed using the information of both relevant FITS keywords stored in the image header and statistical measurements on the image data, (ii) reduction of calibration images (bias, flat-imaging) and generation of combined master calibrations as a function of the modified Julian Date, (iii) basic reduction (bias subtraction, flat fielding of the images, astrometric calibration) of individual science images, making use of the corresponding master calibrations. This software package, although designed to be run on CAFOS images, does allow the inclusion of additional observing modes and instruments.

The master calibration files are generated for each night within a given time span. In this way, science images are reduced with the master calibration files closest in time to the observations. The combination of calibration images also takes into account the signature of the images as described by the following keywords: \verb|CCDNAME|, \verb|NAXIS1|, \verb|NAXIS1|, \verb|DATASEC|, \verb|CCDBINX|, \verb|CCDBINY|.

All the processed images using {\tt Filabres} are available from the Calar Alto archive under the label {\it Advanced Data Products}.

\section{Astrometric calibration}\label{sec:astrometric_calibration}

In this Section, we describe the astrometric calibration carried out with {\tt Filabres} on step (iii). It is performed with the help of additional software tools provided by {\it Astrometry.net}\footnote{\url{http://astrometry.net/doc/readme.html}} and by {\it AstrOmatic.net}\footnote{\url{https://www.astromatic.net/}} using proper motion corrected {\it Gaia} DR2 data \citep{ Gaia2016,GaiaBrown}. {\it Astrometry.net} provides an initial astrometric calibration refined afterwards with the {\it AstrOmatic.net} software, which includes the use of {\tt SCAMP} \citep{scamp}.

Of the 42\,173 images in the Calar Alto archive, 18\,270 have no astrometric reduction at all, 1\,050 do only have the initial astrometric calibration performed with {\it Astrometry.net} tool and 22\,853 images have the additional {\it AstrOmatic.net} solution as well. Therefore, 57\% (1\,050+22\,853) of the images in the archive present an astrometric calibration.

The median of the mean errors in the initial astrometric calibration with the {\it Astrometry.net} tools in the 22\,853 images for which both astrometric solutions were computed is 0.38\,arcsec. This value downs to 0.04\,arcsec after running the {\it AstrOmatic.net} software.

\begin{figure*}
\centering
\includegraphics[width=0.48\hsize, trim=150 0 150 25, clip]{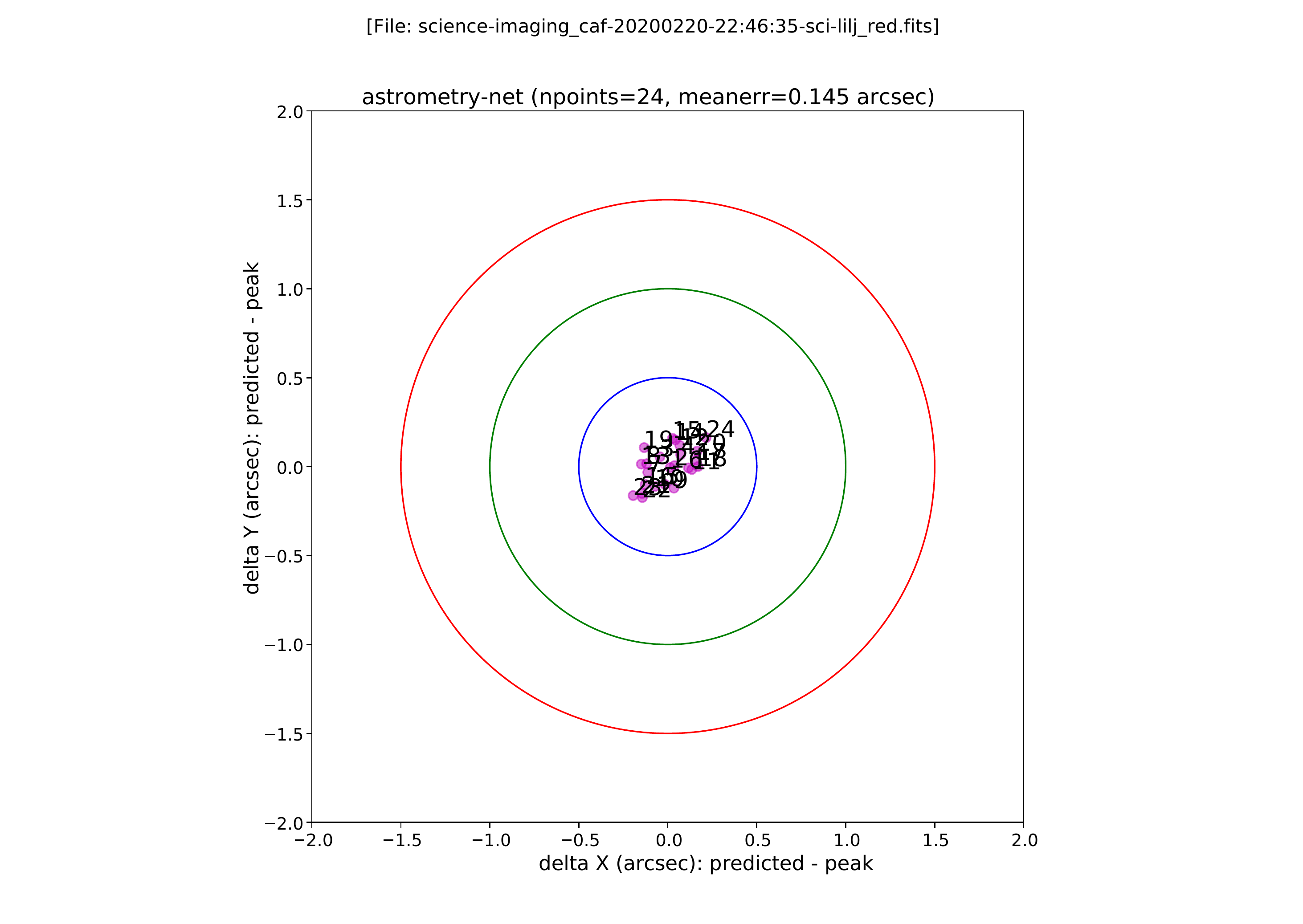}
\includegraphics[width=0.48\hsize, trim=150 0 150 25, clip]{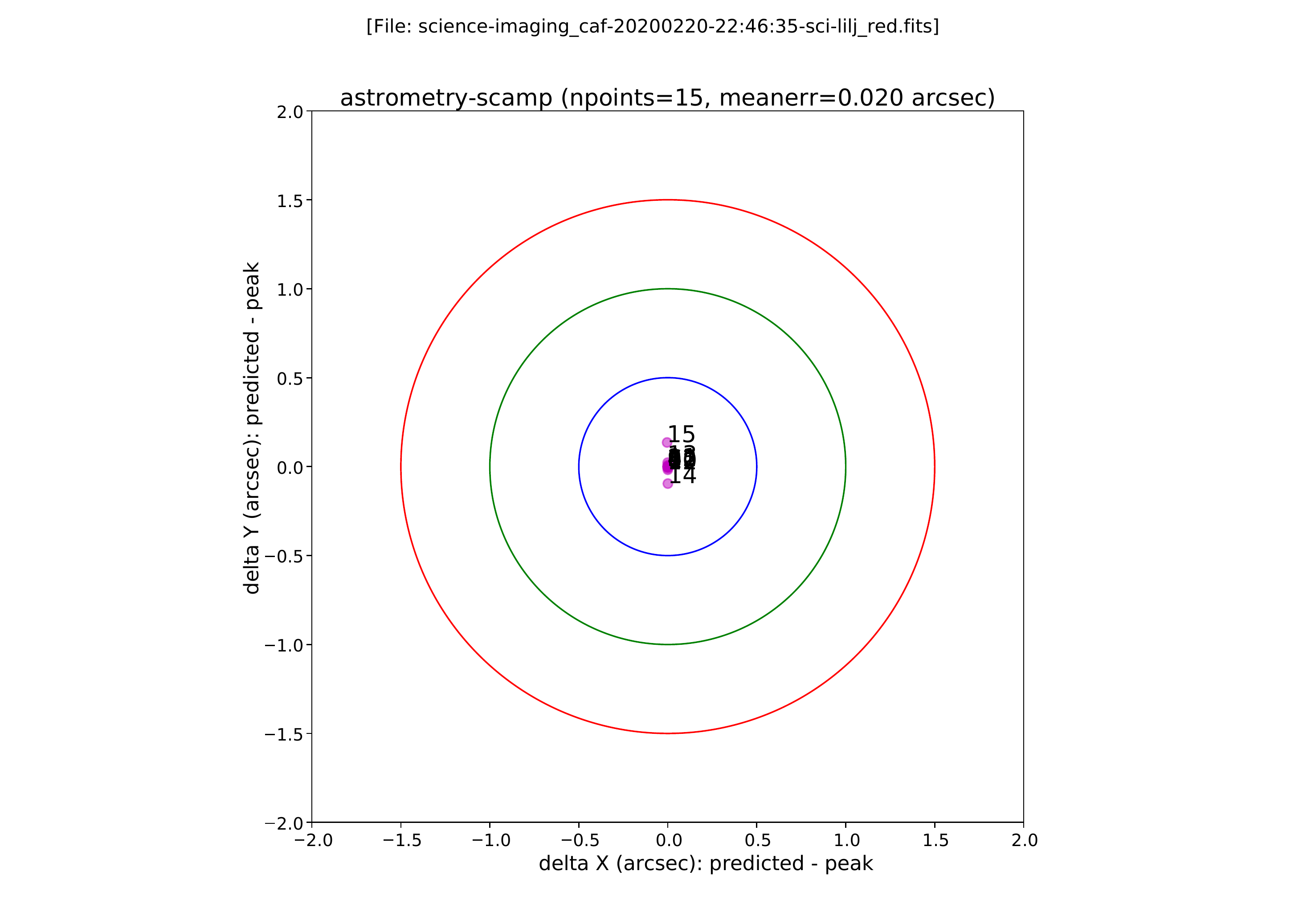}
\cprotect\caption{Differences between the predicted location of the sources and the peak positions using the astrometric solution found with {\it Astrometry.net} (left) and {\it Astrometry.net} + {\it AstrOmatic.net} (right) softwares. Each source is labelled with a number. }
\label{fig:astrometry}
\end{figure*}

Regarding the sample of 6\,616 images observed in the Sloan bands, 6\,132 present both astrometric solutions, with median values of the mean errors of 0.18\,arcsec after the initial calibration and 0.03\,arcsec after the second one.
Figure~\ref{fig:astrometry} compares the differences between the predicted location of the sources and the peak positions found in the images with the {\it Astrometry.net} and {\it AstrOmatic.net} astrometric solutions.

\section{Sources}\label{sec:sources}

\subsection{Source extraction} \label{sec:extraction}

We developed the CAFOS Photometry Calibrator\footnote{\url{https://cafos-photometry-calibrator.readthedocs.io/en/latest/}} ({\tt CFC}), a pipeline designed for (i) measuring the instrumental photometry of sources in CAFOS images, (ii) performing a selection of theses sources using different filtering criteria and (iii) carrying out the photometric calibration.
 {\tt CFC} can be run directly over the {\tt Filabres} structure of folders to use as an input the astrometrically calibrated images provided by the software. It can be run as well over independently reduced and astrometrically calibrated images. In both cases, it is possible to include an additional csv configuration file containing the CAHA numerical identifier, the filter of observation, the name of the file or the path to the images. If the filter field is empty or the pipeline does not identify it, it proceeds to look for it in the image header. By default, it only processes images observed in the Sloan $griz$ bands. All other images will be reported in a log file.

{\tt CFC} performs a step by step process to identify the image sources and estimate their instrumental parameters using {\sc SExtractor} \citep{1996A&AS..117..393B} and {\sc PSFEx} \citep{2013ascl.soft01001B}. The method is based on a first iteration of {\sc SExtractor}, whose output catalogue is used by {\sc PSFEx} to estimate the PSF model of the image sources. The PSF model is then used in a second iteration of {\sc SExtractor} to fit every source using standard PSF-fitting.

We kept sources above 1.5$\sigma$ of the local background and with less than 50\,000 counts as an approximate value for saturation. 
Remaining saturated sources will be removed as described in Section~\ref{sec:calibration}, since they are subject to larger uncertainties in their astrometry and photometry.

We measured \verb|MAG_PSF| photometry, suitable for point-like detections, and we extracted some morphometric parameters like the Full-width at half maximum (FWHM), the flux radius (defined as the circular aperture radius enclosing half the total flux), the signal-to-noise ratio, the elongation and the ellipticity.
We also included the \verb|SPREAD_MODEL| parameter, a star/galaxy estimator based on model fitting. This parameter will be useful for creating our point-like catalogue.
Besides, we add the minimum and maximum x and y coordinates among detected pixels (\verb|XMIN_IMAGE|, \verb|XMAX_IMAGE|, \verb|YMIN_IMAGE|, \verb|YMAX_IMAGE|).

The {\sc SExtractor} and {\sc PSFEx} configuration files used to produce the catalogues of sources are given in Tables~\ref{tab.def_sex} and \ref{tab.def_psfex}, and the first output of the pipeline would be the {\sc SExtractor} catalogues with the columns listed in Table~\ref{tab.def_params}.

\subsection{Source selection}

Aiming at removing spurious detections from the catalogue, we discarded sources with \verb|FLUX_MAX|, \verb|FLUX_RADIUS|, \verb|FWHM_IMAGE| and \verb|FLUX_PSF| smaller than or equal to zero, \verb|MAG_PSF| and \verb|MAGERR_PSF| equal to 99 (meaning that {\sc SExtractor} PSF fit did not converge), \verb|MAGERR_PSF| equal to zero or larger than 1\,mag, and \verb|FLAGS_WEIGHT| (weighted extraction flag related to the presence of close neighbours bright enough to significantly bias the photometry, bad pixels, blended objects, saturated pixels or other features) equal to two. We also neglected sources with \verb|SNR_WIN| smaller than five or equal to 1e30, which has no physical meaning. The limit in \verb|SNR_WIN| at five is set to avoid sources with very poor photometry, while the later is related to a bad extraction of the source.

\begin{figure}
\centering
\includegraphics[width=0.48\textwidth]{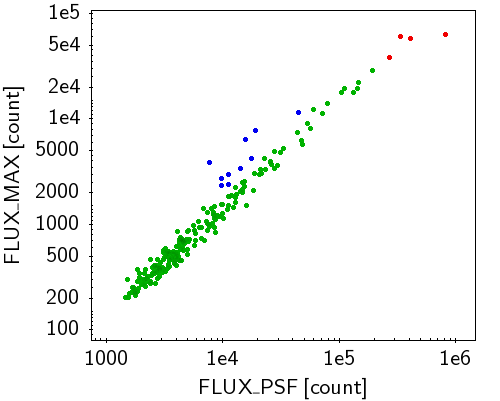}
\cprotect\caption{Flux at the peak (\verb|FLUX_MAX|) vs. integrated flux (\verb|FLUX_PSF|) in logarithmic scale for detections in one image. Spurious, saturated and valid detections are represented in blue, red and green, respectively.}
\label{fig:fluxratio}
\end{figure}

As mentioned before, we set a saturation level when extracting sources in our images at 50\,000 counts. However, some saturated detections could remain unidentified. We attempt to remove them together with the resting spurious detections such as bad pixels, cosmic rays and artifacts with a complementary procedure similar to that followed in \cite{CC_osiris}. It accounts for the linear relation between the flux at the peak of the distribution (\verb|FLUX_MAX|) and the integrated flux (\verb|FLUX_PSF|), as shown in Figure~\ref{fig:fluxratio}. It is a two steps procedure that can summarized as follows: In a first place and in order to discard spurious detections in each image, we kept detections which \verb|FLUX_MAX/FLUX_PSF| ratio deviates by less than 2$\sigma$ from the mean value. We recomputed the mean and standard deviation of the flux ratio of the remaining set of sources and kept detections with ratios under the mean value plus 3$\sigma$. Afterwards and aiming at identifying saturated sources, we look for the value of \verb|FLUX_PSF| in each image at which linearity breaks due to saturation with two simultaneous approaches (i) when the slope of the curve \verb|FLUX_MAX| vs. \verb|FLUX_PSF| varies by more than 1\%, or (ii) when the Pearson correlation coefficient $r$ of the linear fit is lower than 0.98. Both of them are evaluated increasing the number of points towards increasing values of \verb|FLUX_PSF|. The approach that provides the lowest \verb|FLUX_MAX| is applied. We then compute the standard deviation of the sample of sources that present fluxes above that cut value and remove from the catalogue all sources with \verb|FLUX_MAX| greater than the cut value minus 1$\sigma$.

\section{Flux calibration}\label{sec:calibration}

We chose the SDSS DR12 \citep{sdssdr12} and APASS DR9 \citep{APASS} surveys as the photometric references to calibrate our catalogue.
For this, the first step is to make the selection of calibration sources. To select them, we first cross-matched the {\sc SExtractor} catalogue of detections of each individual image with SDSS DR12 within 2.0\,arcsec. We kept all counterparts in that radius and then applied the following criteria to define the calibration sources:

\begin{itemize}

 \item SDSS sources must be point-like (class=6) and flagged with q\_mode='+' and mode=1, indicating clean photometry of primary detections.
 \item SDSS sources must be within the detection and saturation limits\footnote{{For details, see }\url{https://www.sdss.org/dr16/imaging/other_info/}}.
 \item {Both {\sc SExtractor} and SDSS magnitude uncertainties must be smaller than 0.2\,mag}.
 \item FWHM, elongation and ellipticity values must be within the median values of the image and twice the interquantile. This serves to morphometrically identify spikes and extended sources and remove them from our CAFOS catalogue.
 
\end{itemize}

For each image, we kept the sources fulfilling the conditions given above and carried out a sigma-clipping linear fit between the CAFOS {\sc SExtractor} and SDSS DR12 calibrated magnitudes.

We performed the photometric calibration only in images with at least six calibrating point-like sources and when the Pearson correlation coefficient ($r$) of the fit is greater than 0.98. When the sky region was not covered by SDSS DR12 observations or the number of calibration sources was lower than six, we performed the calibration with the APASS DR9 photometric catalogue applying the two latest criteria described above and selecting only APASS counterparts with magnitudes in the range $7 < V < 17$ magnitudes. We calibrated 3\,964 images (out of 6\,616),  3\,334 of which were calibrated with SDSS DR12 and 630 with APASS DR9 photometry. A column indicating the reference catalogue  used for the calibration (SDSS or APASS) is included in the final catalogue. Moreover, since the range of magnitudes reached in each image may be larger than the range of magnitudes of the calibration stars, we also include in the final catalogue a calibration flag that indicates whether the calibrated magnitudes are within the interval of magnitudes used for the photometric calibration ({\tt A}), fainter ({\tt B}), or brighter ({\tt C}). The 71\% of the calibrated magnitudes lies within the interval of magnitudes used for the photometric calibration, only 4\% present fainter magnitudes and 25\% show brighter magnitudes than those used for the calibration.

\begin{figure*}
\centering
\includegraphics[width=0.42\textwidth]{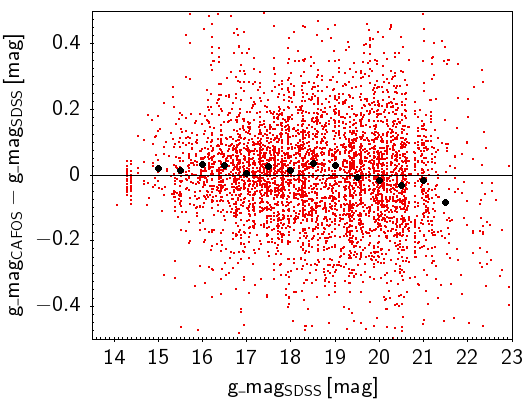}
\includegraphics[width=0.42\textwidth]{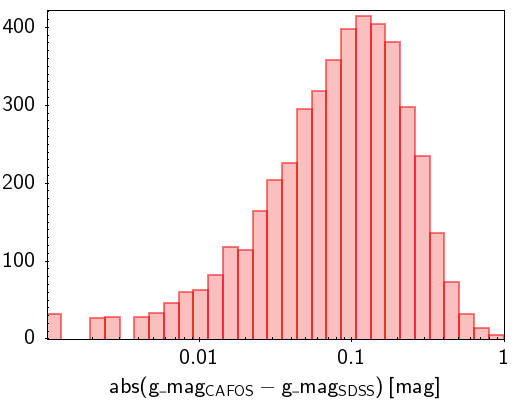}
\includegraphics[width=0.42\textwidth]{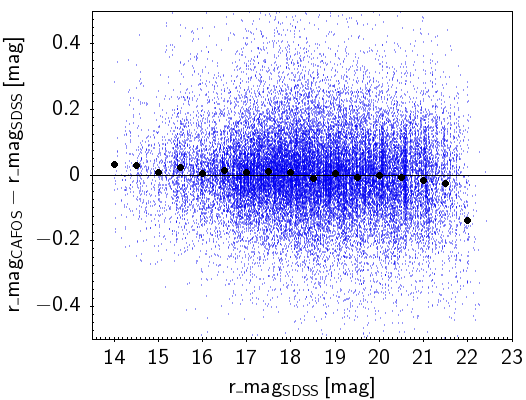}
\includegraphics[width=0.42\textwidth]{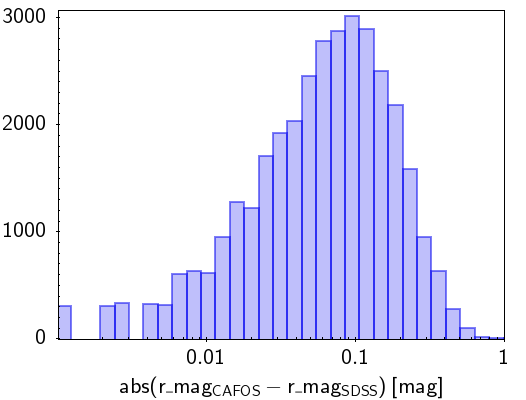}
\includegraphics[width=0.42\textwidth]{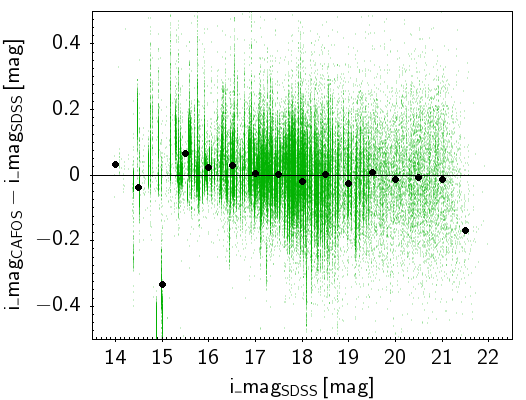}
\includegraphics[width=0.42\textwidth]{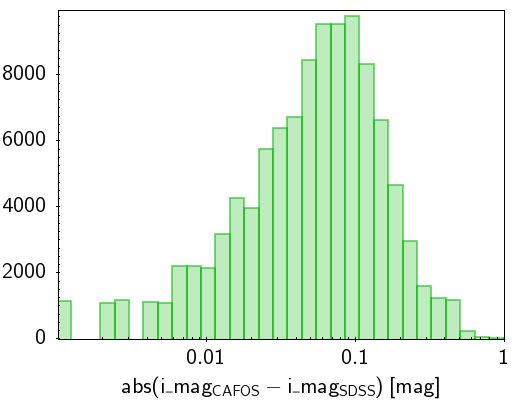}
\includegraphics[width=0.42\textwidth]{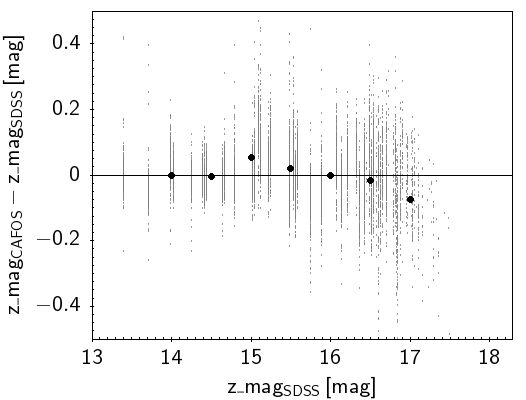}
\includegraphics[width=0.42\textwidth]{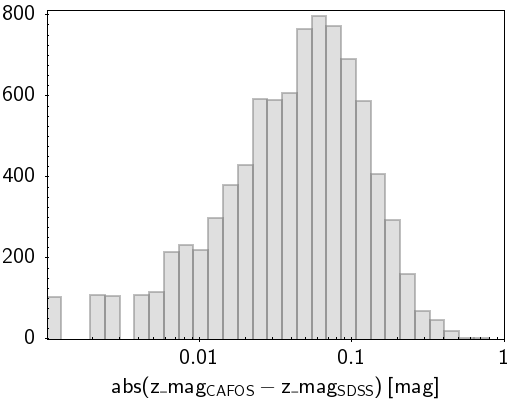}
\caption{Comparison between SDSS DR12 and CAFOS magnitudes of the sources used for the photometric calibration. On the left hand panels, black filled circles represent the average difference of magnitudes in bins of 0.5\,mag. Only bins with ten or more sources are represented. Right hand panels represent the distributions of the absolute values of the magnitude differences.
\label{fig:cal_comparisonSDSS}}
\end{figure*}

\begin{figure*}
\centering
\includegraphics[width=0.45\textwidth]{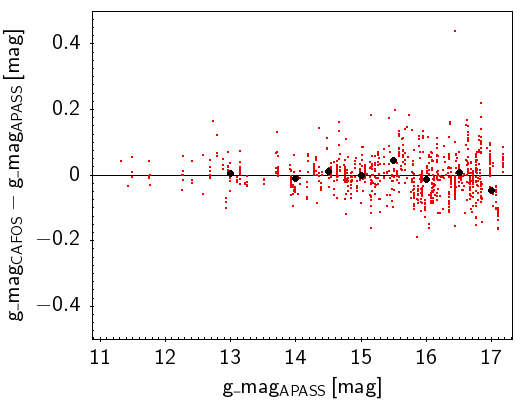}
\includegraphics[width=0.45\textwidth]{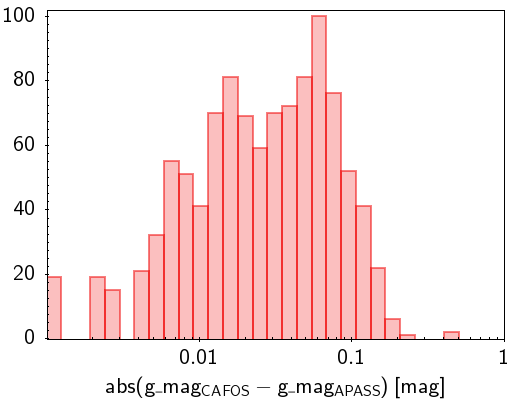}
\includegraphics[width=0.45\textwidth]{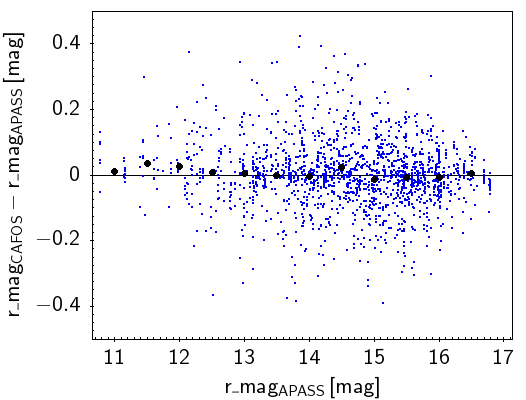}
\includegraphics[width=0.45\textwidth]{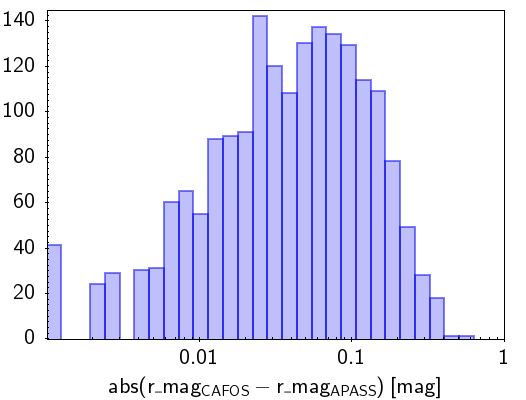}
\includegraphics[width=0.45\textwidth]{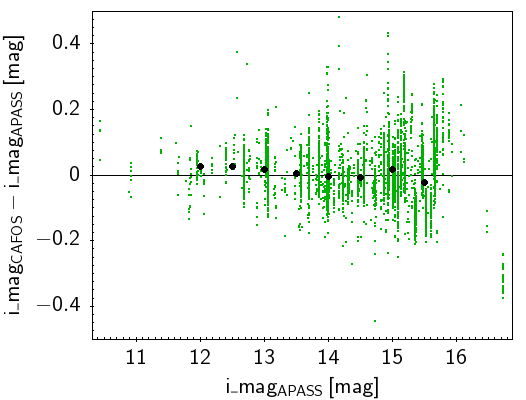}
\includegraphics[width=0.45\textwidth]{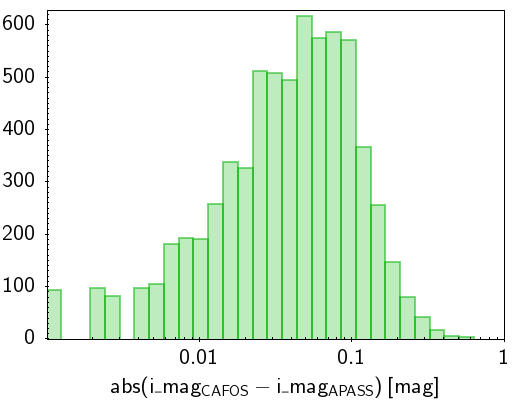}
\caption{Comparison between APASS DR9 and CAFOS magnitudes of the sources used for the photometric calibration. On the left hand panels, black filled circles represent the average difference of magnitudes in bins of 0.5\,mag. Only bins with ten or more sources are represented. Right hand panels represent the distributions of the absolute values of the magnitude differences.
\label{fig:cal_comparisonAPASS}}
\end{figure*}

The calibration residuals of CAFOS magnitudes as a function of the reference magnitudes are shown in Figures~\ref{fig:cal_comparisonSDSS} and \ref{fig:cal_comparisonAPASS}, together with the distributions of the absolute values of the magnitude differences. In the first and third panels, we show the average magnitude differences in bins of 0.5\,mag with more than ten points. In the calibration with SDSS, the average magnitude differences increase at fainter magnitudes (21.5\,mag in the filters $g$ and $i$, 22.0\,mag in $r$ and 17.0\,mag in $z$). The $i$ band presents a large magnitude difference at 15\,mag. A detailed inspection of these detections reveals that the good-quality (q\_mode=`+' and mode=1) SDSS magnitudes are around 0.5\,mag fainter that the secondary measurements (mode=2), which agree better with our photometry, yielding a mean difference of 0.14\,mag instead of -0.41\,mag. With that exception, the averaged differences of magnitudes remain under 0.17 mag in all bands. For the calibration done with APASS, we observe that in the $g$ and $i$ bands the average magnitude difference increases at the edge of the plot at 17\,mag and 15.5\,mag, respectively. In this case, the averaged differences of magnitudes remain under 0.05 mag in all bands. 
Mean magnitude absolute differences (second and fourth panels) vary from 0.06 to 0.12 mag for the comparison with SDSS data and from 0.04 to 0.07 mag in for APASS', depending on the filter. 

\begin{figure}
\centering
\includegraphics[width=0.48\textwidth]{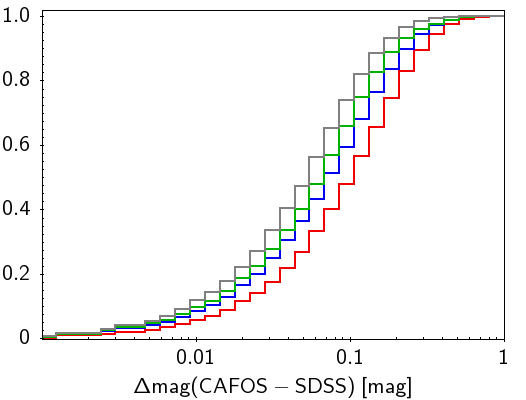}
\includegraphics[width=0.48\textwidth]{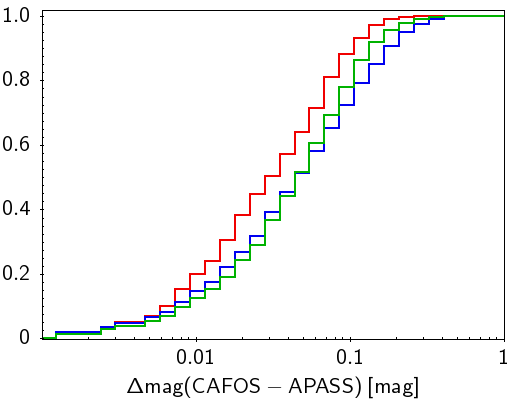}
\caption{Normalized cumulative distribution of absolute magnitude differences between CAFOS and SDSS DR12 (top) or APASS DR9 (down) in logarithmic scale in the $g$ (red), $r$ (blue), $i$ (green) and $z$ (gray) bands. 
\label{fig:cumulativedist}}
\end{figure}

We show the normalized cumulative distributions of the magnitude differences with SDSS DR12 and APASS DR9 in top and bottom panels of Figure~\ref{fig:cumulativedist}, respectively. The photometric scatter in absolute values in 90\% of the sample is under 0.27/0.09, 0.21/0.16, 0.18/0.12 and 0.14 magnitudes in $g$, $r$, $i$, and $z$ bands, respectively, compared to SDSS/APASS.

\section{Image selection}\label{sec:image_selection}

Among the 3\,964 photometrically calibrated images, there are 85 images with no astrometric calibration carried out by {\it AstrOmatic.net} but only by {\it Astrometry.net}. We compared the coordinates of the detections in these images to the astrometry from {\it {\it Gaia}} EDR3 and obtained a mean difference of near 1\,arcsec between coordinates. We considered this value too high to be included in our catalogue and, therefore, these images were discarded. 

Also, it is not uncommon to find in our dataset of images observed with CAFOS blurred images or images out of focus due to bad weather conditions, wrong settings or defocused on purpose.

The image quality has a measurable impact on the {\sc SEXtractor} \verb|FWHM| and the absolute differences between the object position and the windowed position estimate along the x and y axes |\verb|X_IMAGE|$-$\verb|XWIN_IMAGE|| and |\verb|Y_IMAGE|$-$\verb|YWIN_IMAGE|| of the detections. The former is a reflection of the dispersion produced by the telescope and the atmosphere in the photons received from a point source over several pixels in the CCD. The latter is related to the centroiding accuracy of \verb|X_IMAGE| and \verb|Y_IMAGE|, which should be very close to that of PSF-fitting as stated in the documentation of {\sc SExtractor}. Hence, abnormal values (lower or higher than a typical value, as described below) would be associated to low quality images taken under bad weather conditions or with a large fraction of saturated or extended sources.
We therefore used them to assess the quality of an image, together with the mean error of the astrometric calibration provided by {\tt Filabres}.

As a first step, we removed from the catalogue all detections with \verb|FWHM| and \verb|Elongation| that deviate by more than 2$\sigma$ from the median value of each image.
We then identified from visual inspection a sample of 404 good-quality images taken in 2017. These images were used to obtain reference values of the \verb|FWHM|, |\verb|X_IMAGE|-\verb|XWIN_IMAGE||, and |\verb|Y_IMAGE|-\verb|YWIN_IMAGE||. For each image, we obtain the median of each of them. With the set of values of the 404 images, we got three upper limits defined as the median values plus 3$\sigma$ for the \verb|FWHM|, |\verb|X_IMAGE|-\verb|XWIN_IMAGE||, and |\verb|Y_IMAGE|-\verb|YWIN_IMAGE|: 3.96\,arcsec, 0.10\,pix, 0.13\,pix, respectively. We observed that the mean internal uncertainties from {\tt Filabres} take values under 0.09\,arcsec in this set of images. We kept this value as the upper limit for selecting useful images among the photometrically calibrated images.

For each of the remaining 3\,475 calibrated images (3\,964--85--404), we computed the median of the \verb|FWHM|, |\verb|X_IMAGE|-\verb|XWIN_IMAGE|| and |\verb|Y_IMAGE|-\verb|YWIN_IMAGE||. Together with the astrometric error from {\tt Filabres}, we rejected images which median values are greater than the upper limits defined above. We ended up with 2\,338 images from over 55 different nights taken between September 2012 and June 2019. This set contains 218\,782 detections.

By visual inspection, we found that some slightly defocused images still remain. We estimated the fraction of these images to be no larger than 10\%. Nevertheless, the comparison between our calibrated photometry and Sloan's shows differences under 0.1\,mag in average, which is in accordance with what we have presented in Section~\ref{sec:calibration}.

\begin{figure}
\centering
\includegraphics[width=0.45\textwidth]{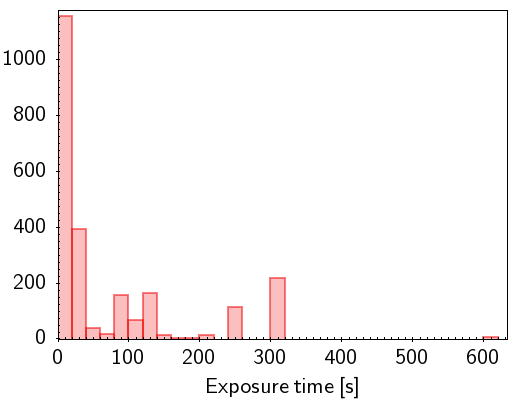}
\caption{Distribution of the exposure times of the 2\,338 images selected for the catalogue.
\label{fig:exptime}}
\end{figure}

The exposure time distribution of the selected images is shown in Figure~\ref{fig:exptime}. In the whole catalogue, exposure time ranges from 0.5 to 600\,s, although in the $z$ band the interval varies from 2.5 to 20\,s. As an average, selected images in all filters have an exposure time of 69\,s and the median of the distribution lies at 20\,s. For very short exposure times, the effect of atmospheric turbulence may be imperfectly time-averaged, meaning the PSF used for extracting fluxes could differ from a smooth Moffat-like function and could also vary over the field of view.
 We observe that a large fraction of the dropped images (85\%) have exposure times under 40\,s (10\,s as a median), which agrees with the before mentioned effects noticeable at short integration times.
 In addition, we also assessed their impact on the astrometric residuals and found no particular trend either on kept or rejected images.

\section{A point-like source catalogue}\label{sec:pointlike}

For the first data release, our main purpose is to provide a point-like photometric catalogue from the CAFOS images with SDSS filters available at the CAHA archive, i.e., a collection of images taken for different science cases and therefore, showing different observational parameters. The choice of a point-like catalogue alone in this first data release is made by simplicity and motivated by the shallower depth compared for instance to the GTC Osiris DR1 catalogue previously published by this group following a similar procedure \citep{CC_osiris}.

After the cleaning in \verb|FWHM| and elongation described in Section~\ref{sec:image_selection}, we used the \verb|SPREAD_MODEL| parameter to identify point-like sources. This parameter shows values close to zero for point sources, positive for extended sources (galaxies), and negative for detections smaller than the PSF, such as cosmic ray hits.
To define the value of the \verb|SPREAD_MODEL| that separates point sources from extended sources and cosmic rays, we made use of the object type provided by the SIMBAD astronomical database \citep{Wenger00}. There are 9\,568 detections in our clean catalogue with a SIMBAD entry within 2\,arcsec. Of them, 5\,795 are classified as ''stars'' and 1\,415 as ''galaxies'' (we excluded any other classification). We observed two close peaks in the distributions of the \verb|SPREAD_MODEL| of these two sets with mean values of $0.008 \pm 0.024$ (stars) and $0.022 \pm 0.018$ (galaxies). The fraction of detections in the negative wings of the distributions is negligible.
After different tests, we defined a detection as point-like if its associated \verb|SPREAD_MODEL| lies in the interval [$-$0.011, 0.013]. This minimizes the contamination of galaxies in the sample and, at the same time, maximizes the completeness of stars. Detections with values below or above that range will be considered as artifacts and extended, respectively, and will be removed from the catalogue. 

To assess the goodness of this approach, we compared our classification with the object type provided by SDSS DR12 under the $class$ column. The completeness of point-like sources reached 89\%, while the contamination is of 15\% in this interval.

We set up different limits in the \verb|FWHM| and ellipticity, and refine the lower limit of the \verb|SPREAD_MODEL|, aiming at removing detections that present anomalous values probably related to a wrong extraction of the sources but also to contamination in the catalogue.
We established an upper limit in the \verb|FWHM| at 3.9\,arcsec from the median and three times the MAD (median absolute deviation) values of the whole catalogue of point-like sources. The \verb|FWHM| lower limit is naturally set by the first centile at 0.8\,arcsec. From the resulting catalogue with \verb|FWHM| between 0.8 and 3.9\,arcsec, we eliminate sources with \verb|SPREAD_MODEL| < $-$0.008 and ellipticities larger than 0.4, corresponding to the first and 99th centiles, respectively.
As a result, the catalogue contains 139\,337 point-like detections from 2\,338 images distributed over 54 different nights.

Detections were grouped into sources by carrying out with STILTS \citep{Taylor06} an internal match of the whole catalogue (this is, regardless the photometric band) within 0.5\,arcsec. This value is a good compromise between completeness and reliability. Larger values may produce associations among unrelated detections. Smaller values could miss faint detections, which typically show large centroiding errors or sources with high proper motions. This way, the 139\,337 detections were grouped in 21\,985 different sources, each of which presents a numerical identifier in the catalogue.
Note that the absence of a source in the catalogue does not imply that it has not been detected by CAFOS but that it might not has passed the quality tests. We thus address him/her to the Calar Alto archive to obtain its photometry from the ready-to-use science images.

Table~\ref{tab.cat-info} lists, for each filter, the number of images and detections contained in the catalogue while Table~\ref{tab.bands-sources} summarizes the number of sources that have been detected in one, two, three or the four bands, regardless which bands they are. Almost 84\% of the sources in the catalogue have been detected in just one band and none has been detected in the four bands. A detailed description of each column in our detection catalogue is given in Table~\ref{tab.catalogue_description}.

\begin{table}
 \centering
 \caption {Number of detections per filter in the catalogue. }
 \label{tab.cat-info}
 \begin{tabular}{c c c}
 \hline \hline
 \noalign{\smallskip}
Filter & 	Number of & 	Number of	 \\
 & images &	detections					\\

 \noalign{\smallskip}
 \hline
 \noalign{\smallskip}
 \noalign{\smallskip}
Sloan g & 169 & 8\,695	 	\\		
Sloan r & 529 & 52\,244		 \\	
 Sloan i & 1\,184 & 69\,759	 \\
Sloan z 	& 456 &	8\,637	 \\
 \noalign{\smallskip} 
 \hline
 \end{tabular}
\end{table}

\begin{table}
 \centering
 \caption {Number of sources detected in one to four bands.} 
 \label{tab.bands-sources}
 \begin{tabular}{c c }
 \hline \hline
 \noalign{\smallskip}
Number of & Number of 	 \\
 bands  & sources		\\

 \noalign{\smallskip}
 \hline
 \noalign{\smallskip}
 \noalign{\smallskip}
4 & 0 (0\%)\\			
3 & 1\,250 (5.7\%)\\	
2 & 2\,356 (10.7\%)\\
1 & 18\,379 (83.6\%)\\
 \noalign{\smallskip} 
 \hline
 \end{tabular}
\end{table}

The mean internal astrometric accuracy in the catalogue is 0.05\,arcsec computed as the quadratic sum of the uncertainties in right ascension and declination as obtained from the {\sc SExtractor} parameters \verb|ERRX2_WORLD| and \verb|ERRX2_WORLD|. The comparison with {\it Gaia} EDR3 within 2\,arcsec after correcting for proper motions provides a median position distance of $0.08 \pm 0.04$\,arcsec, which we consider to be a more realistic assessment of the astrometric accuracy of the catalogue.

The magnitude coverage in the catalogue from the faint to the bright end ranges between 10.6-20.6\,mag in $g$, 10.7-21.0\,mag in $r$, 10.8-20.7\,mag in $i$ and 9.4-16.6\,mag in $z$. For each filter, the brightest magnitude corresponds to the minimum value of the magnitudes in the catalogue. The magnitudes at the faint edge are set by the 90th percentile. The magnitude limits in the CAFOS catalogue are between 1.3 and 2.5 magnitudes brighter than SDSS's in the $gri$ bands, and 4.1\,mag brighter in the $z$ band. The large difference in the $z$ filter is explained by the exposure times of the images. While in the $gri$ filters exposure times are between 55 and 136 seconds as an average, the average exposure time in the $z$ filter is 6\,s.

\begin{figure}
\centering
\includegraphics[width=0.42\textwidth]{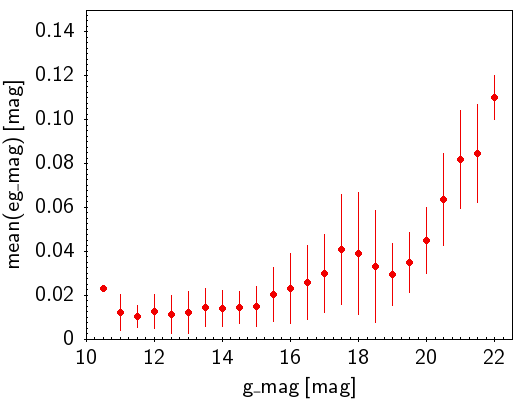}
\includegraphics[width=0.42\textwidth]{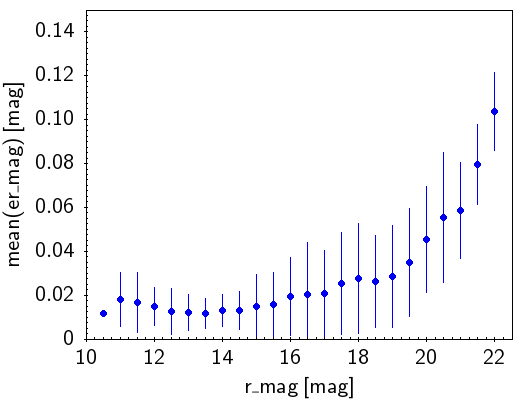}
\includegraphics[width=0.42\textwidth]{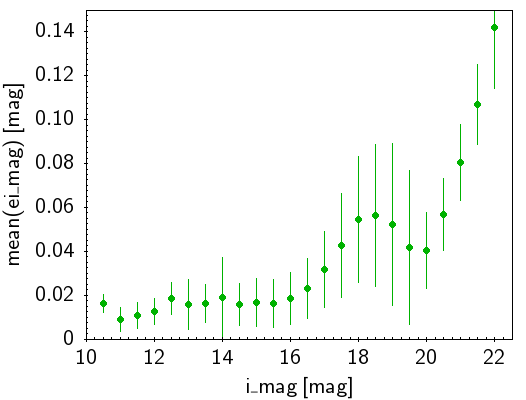}
\includegraphics[width=0.42\textwidth]{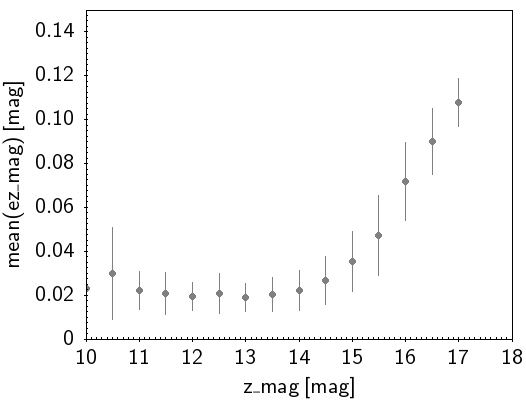}
\cprotect\caption{Internal photometric errors versus magnitudes in the four bands.
Bullets are the average values of the magnitude uncertainties in bin sizes of 0.5\,mag and the error bars show the standard deviation.}
\label{fig:sensitivity}
\end{figure}

To illustrate the sensitivity reached in the catalogue in the four filters, we represent in Figure~\ref{fig:sensitivity} the estimated photometric uncertainties averaged in bins of 0.5 mag, as a function of magnitude. Magnitude uncertainties increase towards fainter magnitudes. Mean magnitude uncertainties are entirely under 0.16\,mag in all bands and the mean internal photometric accuracy of the catalogue is 0.04\,mag. The comparison with SDSS photometry presented in Section~\ref{sec:calibration} yields a more realistic averaged accuracy of 0.09\,mag.

\subsection {Catalogue quality assessment}

\subsubsection{Internal photometric precision}

\begin{figure}
\centering
\includegraphics[width=0.42\textwidth]{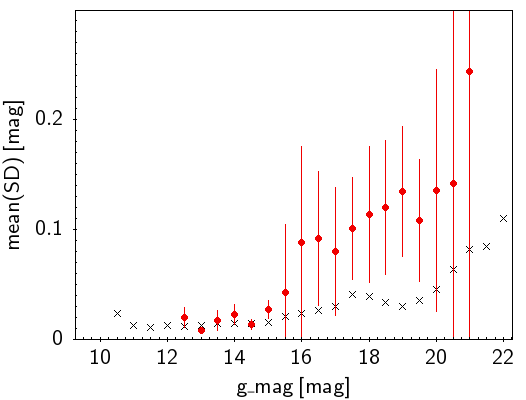}
\includegraphics[width=0.42\textwidth]{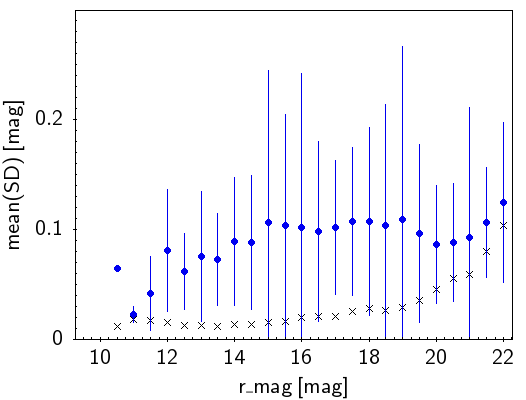}
\includegraphics[width=0.42\textwidth]{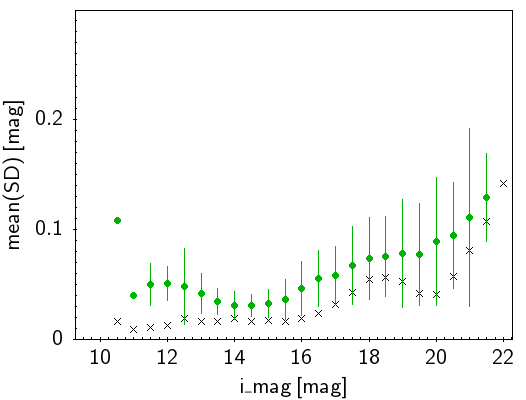}
\includegraphics[width=0.42\textwidth]{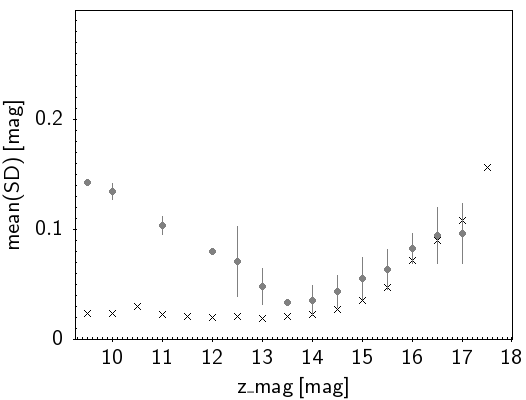}
\caption{Photometric repeatability as a function of magnitudes for sources observed more than five times.
Bullets are the averaged values of the standard deviations of magnitudes of repeated sources in bin sizes of 0.5\,mag and the error bars show their standard deviation.
Mean values of the average photometric errors shown in Figure~\ref{fig:sensitivity} are displayed as black crosses.
\label{fig:stdmag_mag}}
\end{figure}

The comparison of the magnitude of sources observed more than five times in the same filter allows us to estimate the internal photometric precision of the catalogue. We show in Figure~\ref{fig:stdmag_mag} the relation between their standard deviations and magnitudes.
The averaged photometric errors of Figure~\ref{fig:sensitivity} are also represented in the above mentioned figure.

The dispersion in the standard deviation of magnitudes increases towards fainter magnitudes in the $gri$ bands. In the $r$ filter this variation looks flatter and also presents large standard deviations. In the $z$ filter the scatter at bright magnitudes decreases up to 13.5\,mag, where it increases again. The variation of magnitudes of repeated sources is overall over the averaged photometric errors from Figure~\ref{fig:sensitivity}. The largest differences in $g$ in the range between 16 and 21\,mag are partially within the defined error bars, and in $r$ the differences are almost entirely within the error bars. The $i$ band photometry is in good agreement with the mean accuracy of the catalogue. In $z$ band, the discrepancy between magnitude variations and the averaged photometric errors is large at bright magnitudes and up to 13.5\,mag, from where it behaves accordingly. There seems to be no relation between the observed differences in the $g$ and $r$ bands and the exposure times of the images or the number of sources used for the statistics that could explain the larger deviations noted in these bands compared to $i$ and $z$. Revisiting the calibration parameters does not reveal any inconsistency among filters and neither preferred sky positions nor particular epoch of the year of the observations. Other factors that are not taken into account and could affect the photometry could be related to the weather conditions and phase and location of the Moon, for instance. We recall that our sample of data has not been obtained under a systematic configuration but under a diversity of observing configurations adjusted to the different scientific purposes. Among them are variability studies (i.e., transits, stellar spots, pulsations...). The presence of variables in the catalogue is addressed in Section~\ref{sec:variability}.

Despite those apparently larger dispersions in the bluer filters, the averaged standard deviation of sources detected more than five times in each filter is of 0.08\,mag. It is quite consistent with the averaged magnitude differences between CAFOS and SDSS photometry of 0.09\,mag obtained in Section~\ref{sec:calibration}.

\subsubsection{Colour, binning mode and pixel position dependence}\label{sec.colour}

The comparison of the colour differences between CAFOS and SDSS or APASS as a function of the calibrated magnitudes allows us to assess the effect of a colour term in the photometric calibration. As expected, since our images have been observed with the same set of filters used by SDSS or APASS, there is no colour effect.

\begin{figure*}
\centering
\includegraphics[width=0.41\textwidth]{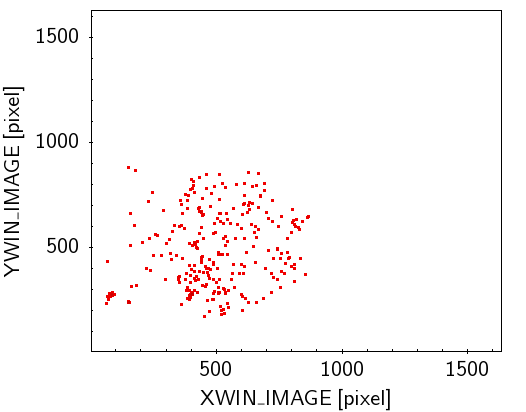}
\includegraphics[width=0.41\textwidth]{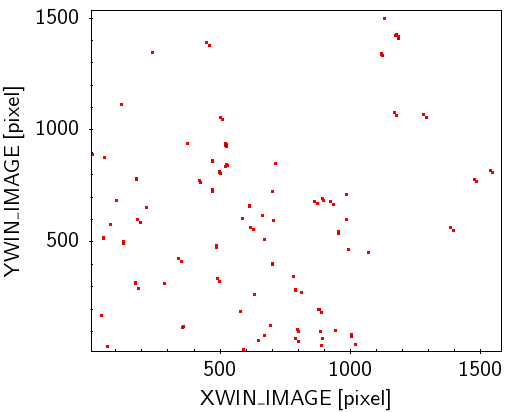}
\includegraphics[width=0.41\textwidth]{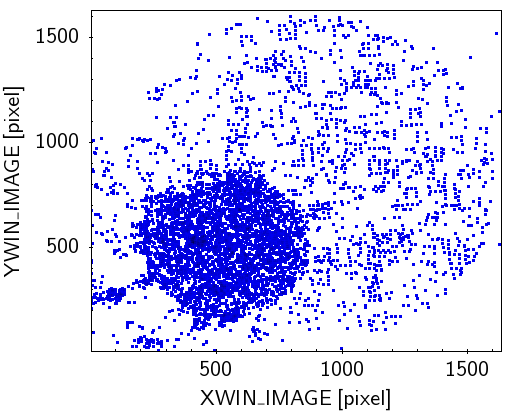}
\includegraphics[width=0.41\textwidth]{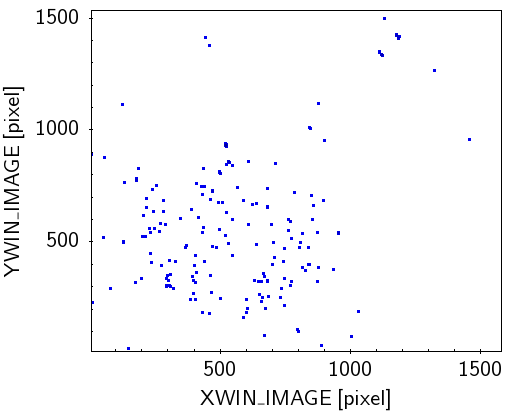}
\includegraphics[width=0.41\textwidth]{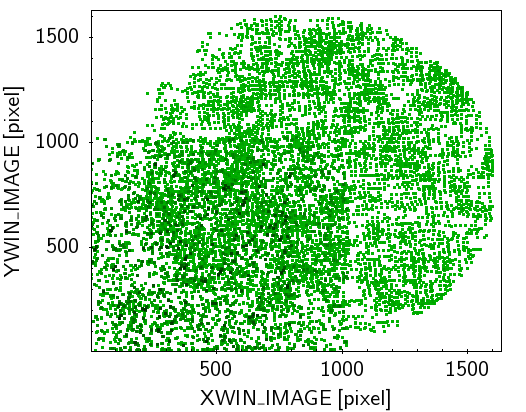}
\includegraphics[width=0.41\textwidth]{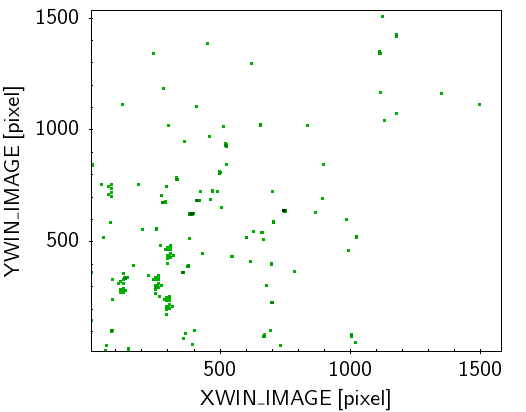}
\includegraphics[width=0.41\textwidth,left=14.5cm]{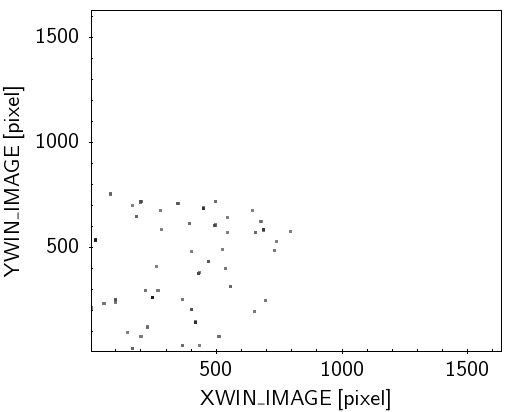}
\caption{CCD $xy$ position of sources with SDSS (left) or APASS (right) good-quality photometry and absolute differences larger than 0.27/0.09, 0.21/0.16, 0.18/0.12 and 0.14 magnitudes in $g$ (red), $r$ (blue), $i$ (green), and $z$ (gray) in SDSS/APASS, respectively.
\label{fig:xy_dep}}
\end{figure*}

We also verify whether there exists a relation of the observed magnitude differences with the pixel position of the sources in the CCD. We show in Figure~\ref{fig:xy_dep} the $xy$ windowed pixel positions from {\sc SEXtractor} of good-quality sources in SDSS DR12 and APASS DR9 that display absolute differences over the mean values in each filter presented in Section~\ref{sec:calibration} (0.27/0.09, 0.21/0.16, 0.18/0.12 and 0.14 magnitudes in $g$, $r$, $i$, and $z$ in SDSS/APASS, respectively). The different observing modes of CAFOS are clearly seen.
We observe some overdensities in the $r$ band in images where SDSS is the reference catalogue for photometric calibration at different positions of the CCD. For example, those located around the pixel positions (100,275), (230,50), (420,530) are clearly identified and correspond to over 170 detections of the same source. A similar behaviour is observed in the $i$ band in images where APASS is the reference catalogue for photometric calibration at (260,310), (300,220) or (730,20), for instance, with more than 35 detections of the same target. Hence, these clumps could not be ascribed to a pixel position dependence but to different measurements of the same object over the same observing run. 

In general, the random configuration of the sources in the CCD reflects that the magnitude differences do not depend on the binning mode of observation nor the pixel position. 

\subsubsection{Variability}\label{sec:variability}

Over 54 nights, the catalogue maps 89 different sky regions, some of which have been intensively surveyed. 

To assess whether the observed differences between magnitudes of repeated objects and the mean accuracy of the catalogue are associated to real variable sources, we analyze the magnitude variations according to the following equation:
\begin{equation}
S = {\rm mag}_{\rm min} + 5 \cdot {\rm e\_mag}_{\rm min} - ({\rm mag}_{\rm max} - 5 \cdot {\rm e\_mag}_{\rm max})
\end{equation}

\noindent where mag$_{\rm min}$, mag$_{\rm max}$, e\_mag$_{\rm min}$ and e\_mag$_{\rm max}$ are the lower and higher magnitudes measured for the same source and their associated errors. We chose 5$\sigma$ to take into account for the underestimated photometric uncertainties in the catalogue.

We crossmatched the whole catalogue with the AAVSO International Variable Star Index VSX \citep{aavso} and obtained 51 different sources that have been observed with CAFOS in at least one filter. We kept the 21 sources observed five or more times in the same filter in our catalogue and obtained an average value of  $S=-0.038 \pm 0.082$\,mag for variable objects. A negative value of $S$ means that the magnitude deviation of the source is outside the reach of ﬁve times the errorbars and cannot be explained by photometric errors. Aiming at minimizing the contamination of sources that could present large scatter in magnitudes due to the different observing conditions, we set $S$ at --0.12\,mag (median -- 1$\sigma$) and looked for variable candidates in our catalogue that fulfill $S<-0.12$\,mag.

\begin{figure}
\centering
\includegraphics[width=0.45\textwidth]{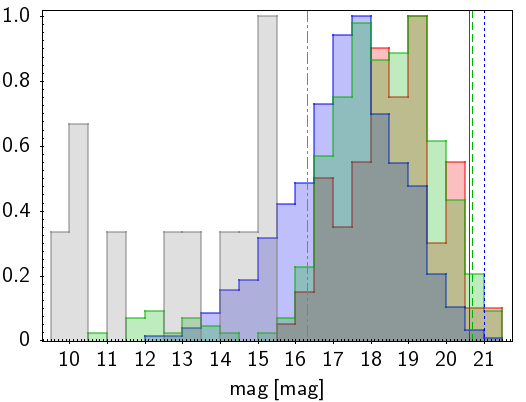}
\caption{Normalized distributions of magnitudes of sources with $S<-0.12$ in the $g$ (red), $r$ (blue), $i$ (green) and $z$ (black) bands. Vertical coloured solid, dashed and dot-dashed lines represent the magnitude detection limits in the catalogue for each filter with the same colour code.
\label{fig:hist_var}}
\end{figure}

\begin{figure*}
\centering
\includegraphics[width=0.33\textwidth]{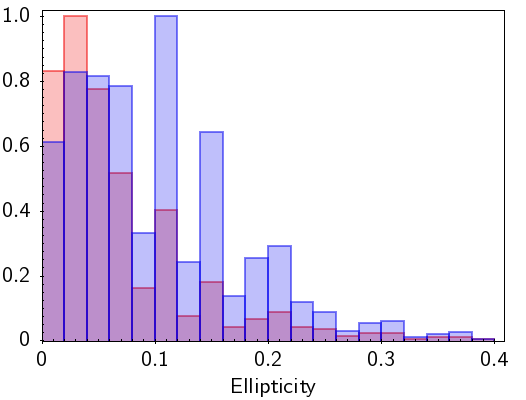}
\includegraphics[width=0.33\textwidth]{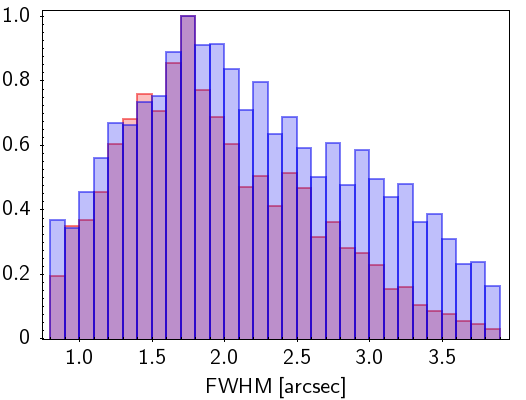}
\includegraphics[width=0.33\textwidth]{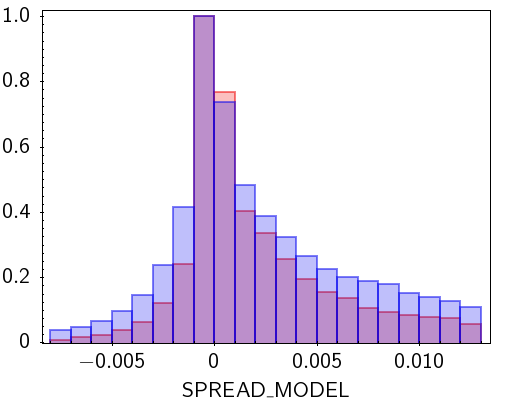}
\caption{Normalized distributions of the ellipticity, FWHM, and SPREAD\_MODEL from left to right of detections repeated more than five times in each filter with $S<-0.12$\,mag (red) and with $S \ge -0.12$\,mag (blue).
\label{fig:snreloellyfwhm}}
\end{figure*}

Between 8 and 33\% of the sources that have been detected at least five times in the same filter have associated a value of $S<-0.12$\,mag, depending on the filter. We show the normalized distribution of magnitudes of these sources in the four filters in Figure~\ref{fig:hist_var}. As a reference, we display in each filter the magnitude detection limits as defined in Section~\ref{sec:pointlike}. The peaks of the distributions are well below these limiting magnitudes. We conclude that the magnitude variations can not be entirely ascribed to faint sources which could present lower quality photometry.
In Figure~\ref{fig:snreloellyfwhm} we compare the ellipticity, \verb|FWHM| and \verb|spread_model| of detections with $S$ over and below --0.12\,mag. 
The distribution of ellipticities, \verb|FWHM| and \verb|spread_model| are similar in detections with $S<-0.12$\,mag (suspected variables) and with $S \ge -0.12$\,mag. This means that potential biases in the morphometric parameters https://www.overleaf.com/project/60d5cdc54d3e31d7eb91088eare responsible for the observed variability.

\begin{figure}
\centering
\includegraphics[width=0.45\textwidth]{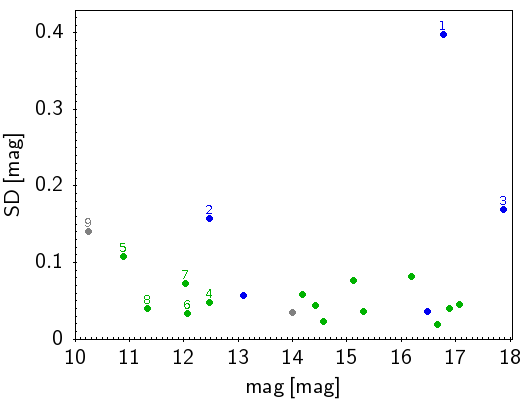}
\caption{Standard deviation of magnitudes of known variable sources compared to their mean magnitudes. Blue, green and gray filled circles stand for r, i and z bands, respectively. Numbers refer to the nine sources with a standard deviation equal to or larger than the difference between the maximum and the minimum magnitude (or the amplitude), as provided in VSX. See the text for discussion.
\label{fig:var_plot}}
\end{figure}

On the other hand, we compared the standard deviation of magnitudes of the 21 sources with counterparts in the VSX catalogue with at least five measurements in the same band, with respect to their mean magnitudes in the corresponding filter (Figure~\ref{fig:var_plot}). We identify nine sources with a standard deviation equal to or larger than the difference between the maximum and the minimum magnitudes (or the amplitude), as provided in VSX. Each of them are labelled from 1 to 9 in Figure~\ref{fig:var_plot}. 

Source \#1 (CRTS J074523.2+460549) is a known eclipsing binary that presents the largest variation in the $r$ band (SD = 0.4\,mag). Sources \#2 (NSV 25135) and \#3 (ZTF J065139.43-002656.9) also show relatively large variations in the $r$ band. The former is catalogued in VSX as a suspected variable that lacks deeper study and the later is defined as a BY Draconis-type. The variation in amplitude estimated from the CAFOS images (0.15\,mag) is in agreement with what is expected from these flaring stars. Also, the chromospheric activity associated to this class of objects could explain the large variations seen in the $r$ band. 
Sources \#4 (WASP-36), \#6 (V1434 Her), and \#8 (V0357 Del) are classified as variable stars in SIMBAD. The moderate activity of source \#7 (HAT-P-12) as described in \citep{Knutson10} might justify the photometric variability. All of them present standard magnitude variations under 0.07\,mag in the $i$ band.
Sources \#5 and \#9 are the same object, named HAT-P-20. It has been detected in the $i$ and $z$ bands in near 140 images. The magnetic activity of the star \citep{Sun17} explains the observed magnitude variations.

Hence, our catalogue might be suitable for detecting variables.
From the previous exercise, we obtain an average standard deviation of 0.2\,mag using the five more scattered sources in Figure~\ref{fig:var_plot} and 1$\sigma$.
We therefore combine the $S<-0.12$ cut with the magnitude standard deviation over 0.2\,mag of sources detected more than five times in the same filter to identify variable candidates in our catalogue.

Out of the 2\,457 unique sources in each filter (note that there can be sources detected in more than one filter) with $S<-0.12$ and five or more detections, 28, 151 and 32 have a standard deviation of magnitudes larger than 0.2\,mag in the $g$, $r$, and $i$ bands, respectively. None in the $z$ satisfies both criteria. Accounting for two targets that fulfill the criteria in more than one filter, there are a total of 209 different sources in the catalogue that are likely photometric variables. Keeping sources with $S<-0.12$ in more than one filter drastically reduces the chances of being photometric variations associated to atmospheric conditions or noise not covered by the photometric uncertainties. They are listed in a separate table in the archive (see Section~\ref{sec:data_access} for details) and at CDS through the VizieR service. A more detailed study including a proper characterization of these sources is beyond the scope of this paper.

\section{Scientific exploitation}\label{sec:science}

To illustrate the science capabilities of the CAFOS DR1 catalogue, we have defined the following three use cases.

\subsection{Identification of asteroids}\label{subsec:ast}

Solar System Objects (SSOs) such as asteroids are frequently found serendipitously crossing the field of view of astronomical observations.  Their fast apparent motion due to their closeness makes challenging their identification.
The {\tt ssos} pipeline \citep{Mahlke2019} has been developed to detect and identify SSOs based mainly on their linear apparent motion in subsequent exposures. The detection and association of the sources are performed with {\sc SExtractor} and {\tt SCAMP}, respectively. SSOs are distinguished from other sources in the images by performing a chain of user-configurable filter algorithms.

Before running the pipeline, we grouped the 6\,616 CAFOS images observed in SDSS filters by sky region and observation run. There are 155 groups with four or more exposures, which is the minimum required for obtaining a reliable detection of a SSO. A total of 6\,055 images are considered for the search. 

\begin{figure}
\centering
\includegraphics[width=0.95\columnwidth]{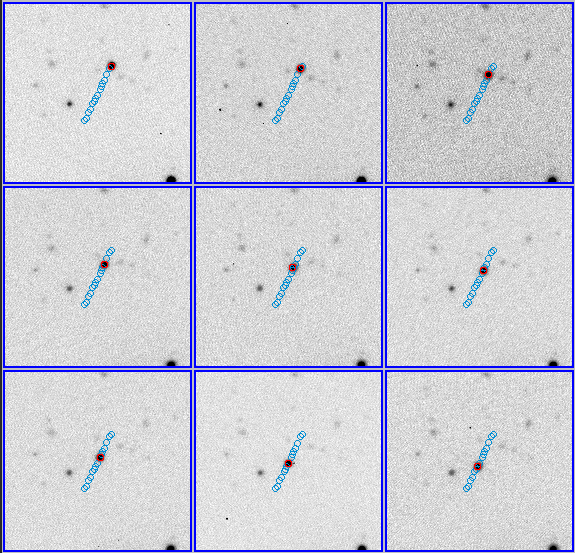}
\caption{Example of linear motion of the Apollo asteroid Sisyphus during nine consecutive images centered at RA=12:39:49.3 and DEC=+52:13:52.0. The temporal coverage of the full set is of near 46 minutes. The epoch increases from left to right and from top to bottom in intervals of 3.3 minutes as an average. The proper motion of the asteroid is 41.5''/h. Blue circles indicate the trajectory of the asteroid in a $\sim$2.6 square arcmin sized field. Red circles indicate the current asteroid position in the image.
\label{fig:traza_asteroide}}
\end{figure}

The {\tt ssos} pipeline identified 20 SSOs with 446 detections in 341 images, 13 of which are already known according to the IMCCE Virtual Observatory Solar System Portal's SkyBoT service\footnote{\url{http://vo.imcce.fr/webservices/skybot/}}. This service computes the ephemerides of SSOs within a given field-of-view and observation epoch, which were cross-matched with the positions of the recovered SSOs in a 10\,arcsec radius.
The number and classification of previously known SSOs detected in our images are summarized in Table~\ref{table:asteroids}.
The remaining seven SSOs are not known, having no registered asteroids with larger position uncertainties within 1\,arcmin.
Astrometry of the 446 detections has been reported to the Minor Planet Center\footnote{\url{https://www.minorplanetcenter.net/}}. Their acceptance proves the goodness of the astrometry in the CAFOS catalogue. An example of an Apollo known asteroid identified in the images is shown in Figure~\ref{fig:traza_asteroide}.

The low number of detections gathered and the short temporal baseline of each detected SSO prevents us from obtaining a spectrophotometric classification or a rotational period from the light curve analysis. Nonetheless, the combination of our observations with other data will help in the characterization of these asteroids.

\begin{table}
        \centering
        \caption {Classification of previously known SSOs detected in the DR1 images. MB stands for Main-Belt. Number of detections in brackets.}
        \label{table:asteroids}
        \begin{tabular}{lllll}
        \hline \hline
        \noalign{\smallskip}
Comet  & Inner MB &Middle MB &Outer MB & Apollo \\
\noalign{\smallskip}
        \hline
1 (63)  & 4 (123) & 1 (12) & 6 (61) & 1 (144) \\
        \noalign{\smallskip}
        \hline
        \end{tabular}
\end{table}

\subsection{Transient candidates}

\begin{figure*}
\centering
\includegraphics[width=0.33\textwidth]{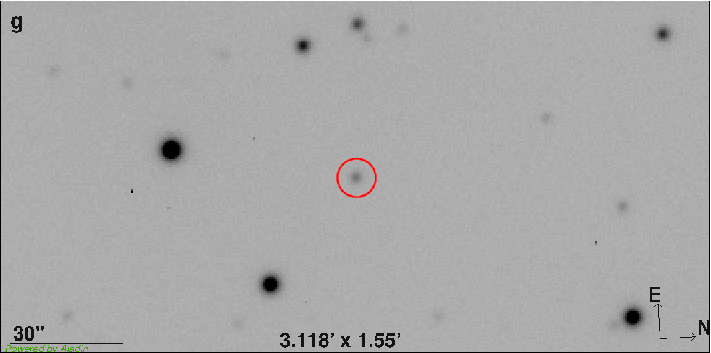}
\includegraphics[width=0.33\textwidth]{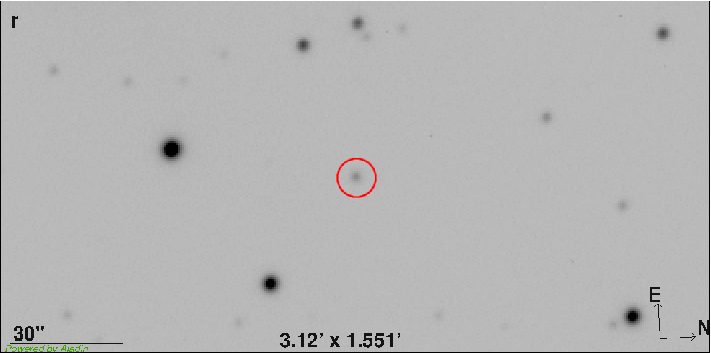}
\includegraphics[width=0.33\textwidth]{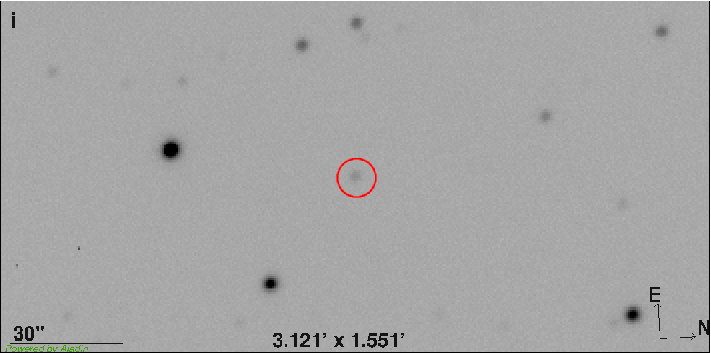}
\caption{Example of a source (GRB 190627A, $i$=18.2 mag), detected in the $g$, $r$ and $i$ bands and with no counterpart in the {\it Gaia} EDR3, SDSS DR12, APASS DR9 or Pan-STARRS DR1 surveys.
\label{fig:346.59-21.2842}}
\end{figure*}

The resulting catalogue provides accurate astrometry and photometry for 139\,337 detections in the optical $griz$ bands. Among those detections, we identify 523 entries corresponding to 516 unique sources that have no counterpart in the {\it Gaia} EDR3, SDSS DR12 or Pan-STARRS DR1 \citep{Kaiser10,Chambers16} optical surveys within 5\,arcsec. All of them lie in the sky region covered by {\it Gaia} EDR3 and Pan-STARRS DR1 and all but 17 in the region covered by SDSS DR12. These surveys are deeper compared to CAFOS and all the 523 entries studied here are brighter than 21.3\,mag, thus, being expected to have been detected in the aforementioned surveys.
They represent the 0.4\% of the CAFOS photometric catalogue. Figure~\ref{fig:346.59-21.2842} shows one of these sources observed in the $g$, $r$, and $i$ filters, as an example.

\begin{figure}
\centering
\includegraphics[width=0.41\textwidth]{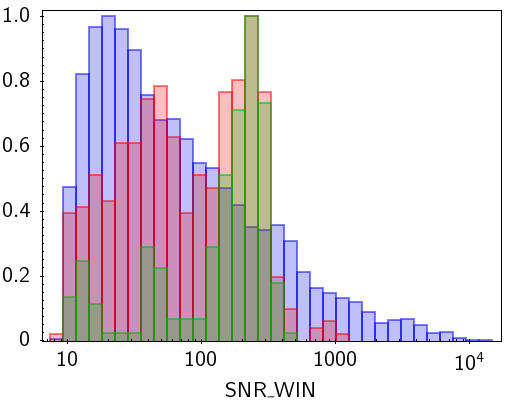}
\includegraphics[width=0.41\textwidth]{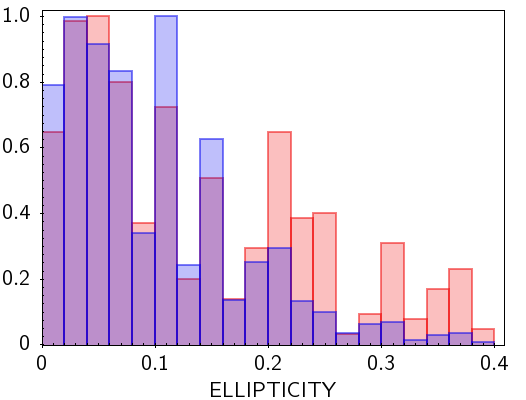}
\includegraphics[width=0.41\textwidth]{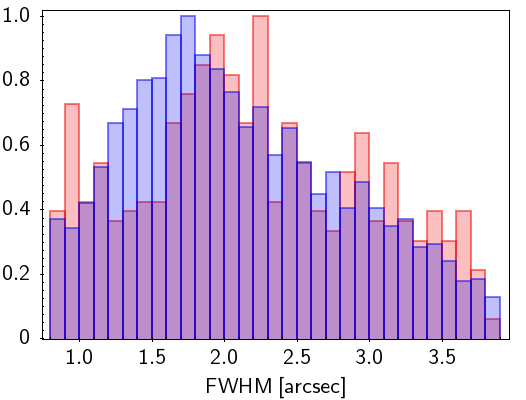}
\includegraphics[width=0.41\textwidth]{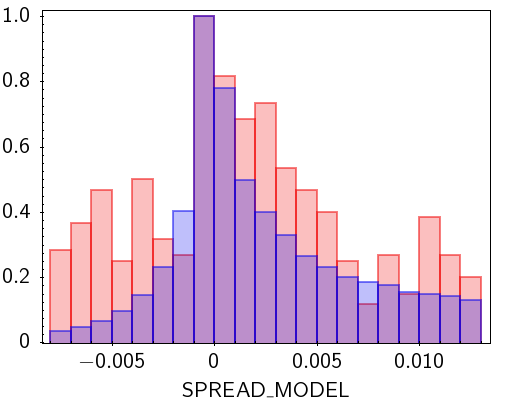}
\caption{Normalized distributions of the signal-to-noise ratio, ellipticity, FWHM, and SPREAD\_MODEL from left to right of the 523 detections with no counterpart in optical surveys (red) and in the whole catalogue (blue). Green on the first panel stands for the SSOs in the sample.
\label{fig:hist_436}}
\end{figure}

The normalized distributions of the signal-to-noise ratio, ellipticity, \verb|FWHM| and \verb|spread_model| of the 523 detections is shown in Figure~\ref{fig:hist_436}. The distributions of the whole catalogue are displayed as well for comparison. We can see how the histogram of the signal-to-noise ratio of the transient candidates shows two peaks, in contrast to the distribution of the catalogue, which presents only one. 
Out of the 523 detections, 212 (40\% of the sample) correspond to asteroids identified in the previous analysis with {\tt ssos} and explain the second peak in the distribution (see the first panel in Figure~\ref{fig:hist_436}). An example of one of them is shown in Figure~\ref{fig:traza_asteroide}.

Out of the remaining 304 sources (516--212), only one has an entry in the SIMBAD astronomical database: GRB 190627A, classified as a gamma-ray burst (Figure~\ref{fig:346.59-21.2842}). We also found 22 counterparts in the Zwicky Transient Facility \citep[ZTF,][]{ztf_Bellm_2018,ztf_Masci_2018}) archive, one in the GALEX GR6+7 AIS \citep{galexgr6+7} catalogue, 22 in UKIDSS LAS9 \citep{UKIDSS2007}, DENIS \citep{DENIS1999,DENIS2000} and/or ALLWISE \citep{allwise}, and 184 counterparts in NEOWISE \citep{neowise}, all of them within 5\,arcsec. All of these counterparts correspond to 198 sources. This is a remarkable result as it confirms the real nature of (at least, a significant fraction of) our transient candidates discarding other options like artifacts or spurious features that may have escaped from the quality control process.

 Sources with NEOWISE counterparts present $W1-W2$ colors between $-3.8$ and 4.3\,mag, regardless the photometric quality in the catalogue. The NEOWISE catalogue also provides the photometry of the associated ALLWISE source, which $W3$ band serves us to identify non extragalactic sources by applying the relation $W1 - W2 > 0.96(W2-W3) -0.96$ derived by \citep{Kirk11}. There are 56 non extragalactic sources in our sample with typical $W1-W2$ colours of red and very red objects (from late K to late T/Y dwarfs and subdwarfs, e.g. Figure~1 in \citealt{Kirk11}). This sources are properly identified in the catalogue.

\begin{figure*}
\centering
\includegraphics[width=0.45\textwidth,trim=0 0 4.4cm 0,clip]{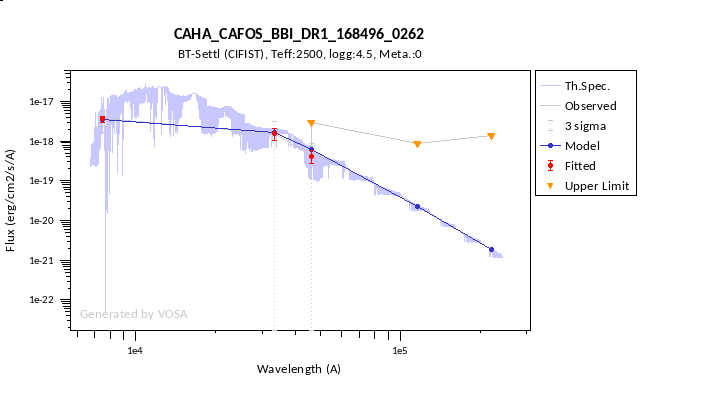}
\includegraphics[width=0.45\textwidth,trim=0 0 4.4cm 0,clip]{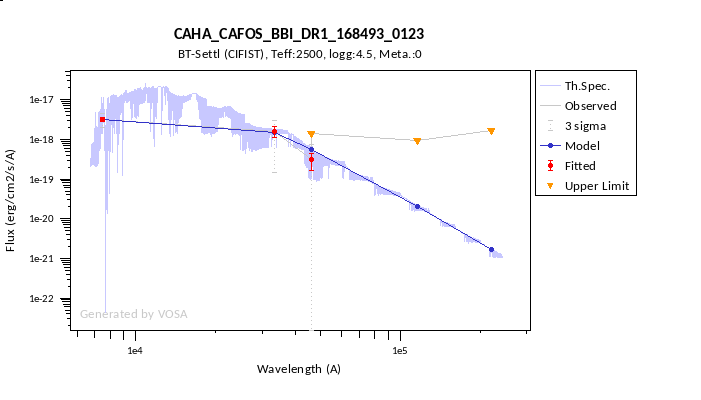}
\includegraphics[width=0.45\textwidth,trim=0 0 4.4cm 0,clip]{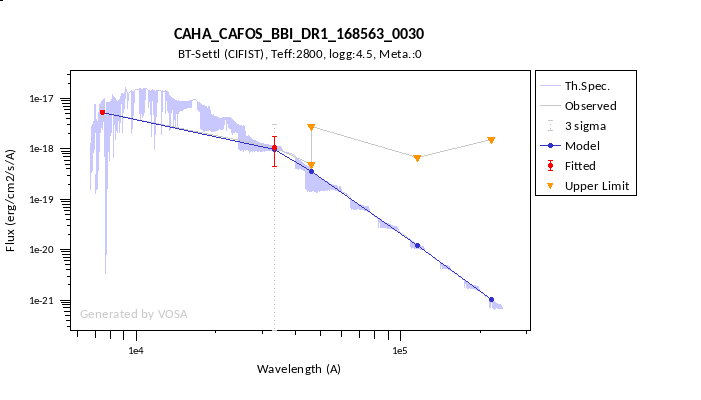}
\includegraphics[width=0.45\textwidth,trim=0 0 4.4cm 0,clip]{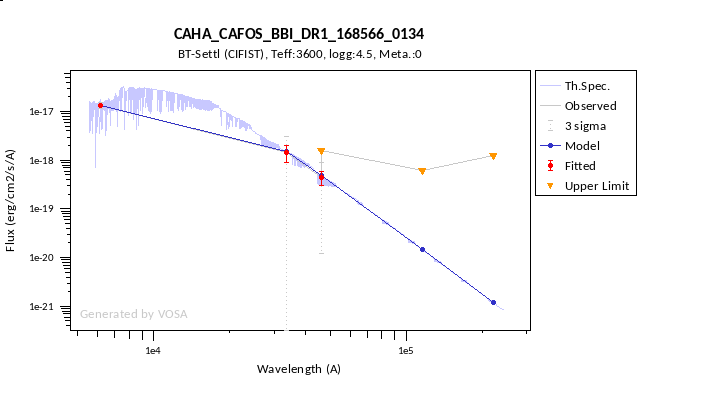}
\includegraphics[width=0.45\textwidth,trim=0 0 4.4cm 0,clip]{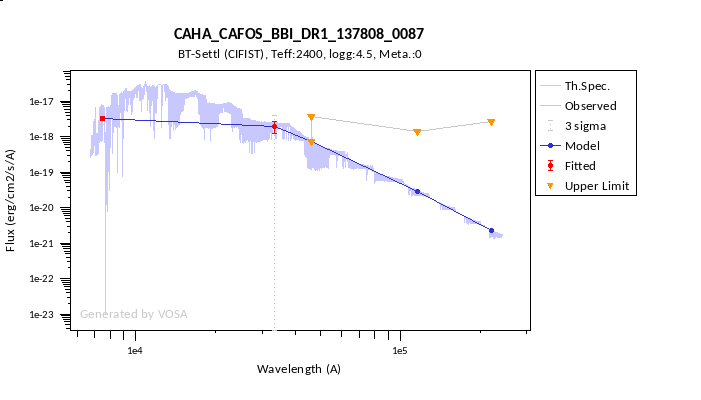}
\includegraphics[width=0.45\textwidth,trim=0 0 4.4cm 0,clip]{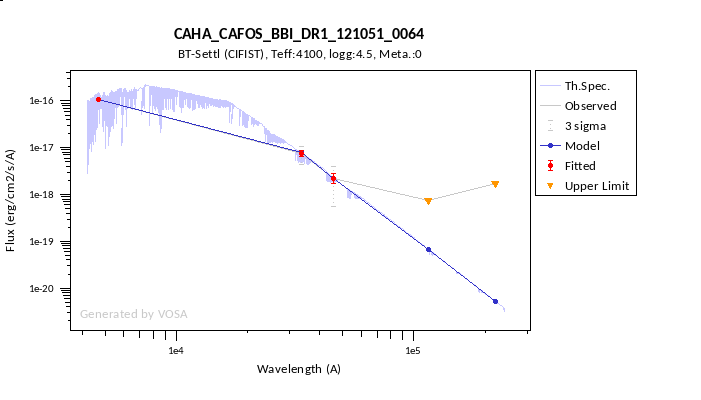}
\caption{Spectral energy distributions of the six low mass stars identified with VOSA with no optical counterpart. The blue spectrum represents the theoretical model that best fits and red circles represent the observed photometry used for the fit. The inverted yellow triangles indicate that the photometric values correspond to upper limits, which generally come from the $W2$ band in WISE or NEOWISE, and $W3$ and $W4$ WISE bands.
\label{fig:vosa}}
\end{figure*}

We then attempt to characterize them by expanding the wavelength coverage using the photometric catalogues available from VOSA \citep{Bayo08}, a Virtual Observatory tool to estimate physical parameters from the spectral energy distribution (SED) fitting to theoretical models. We gather photometry from GALEX-GR6+7, CAFOS, ZTF, DENIS, UKIDSS, WISE and NEOWISE, and perform a $\chi^{2}$ fit using the BT-Settl CIFIST grid of theoretical models \citep{Husser13,Baraffe15} with 1\,200 < $T_{\rm eff}$ [K] < 7\,000, $\log g$ = 4.5 [dex] (sources were assumed to be dwarfs, although the impact of gravity on the model fitting is of second order), and solar metallicity. We obtain effective temperatures between 2\,400 and 4\,100\,K for the six objects listed in Table~\ref{tab.vosa}. The internal uncertainty in the temperatures is 50\,K, estimated as half the grid step of the model, although real temperature uncertainties are definitely  larger since the maximum of the SED is poorly represented. Their corresponding SEDs are displayed in Figure~\ref{fig:vosa}. These temperatures are compatible with a late-type K and five M solar metallicity dwarfs. All of them could thus be flaring objects that might have been observed with CAFOS during the flash events. Nevertheless, spectroscopic follow-up observations would be required for a proper characterization of these objects.

\begin{table}
\small
\setlength{\tabcolsep}{3pt}
\centering
 \caption {Targets of interest characterized with VOSA.}
 \label{tab.vosa}
 \begin{tabular}{lccc}
 \hline \hline
 \noalign{\smallskip}
Detection ID & RA &   DEC &	$T_{\rm eff}$    \\
 &   [hms]   &[dms]  & [K]	\\

 \noalign{\smallskip}
 \hline
 \noalign{\smallskip}
 \noalign{\smallskip}
 
CAHA\_CAFOS\_BBI\_DR1\_168496\_0262   & 01:51:12.13 & +10:46:33.4  &   2\,500    \\ %
CAHA\_CAFOS\_BBI\_DR1\_168493\_0123 	&  01:51:37.11  &  +10:50:14.4&	2\,500    \\ %
CAHA\_CAFOS\_BBI\_DR1\_168563\_0030 &   01:51:51.73 &   +10:40:20.5 &   2\,800  \\
CAHA\_CAFOS\_BBI\_DR1\_168566\_0134  &   07:58:17.13 &   +52:41:39.3 &   3\,600  \\
CAHA\_CAFOS\_BBI\_DR1\_137808\_0087 &   12:15:41.78 &   +02:07:49.2	& 	2\,400 	    \\ %
CAHA\_CAFOS\_BBI\_DR1\_121051\_0064    &  23:12:59.21 &   +11:07:53.8	  & 	4\,100    \\ %

\noalign{\smallskip} 
 \hline
 \end{tabular}
\end{table}

As a result, 105 ($516\!-\!212\!-\!1\!-\!198$) objects observed at different epochs and randomly distributed in the sky remain without further information. Visual inspection reveals that near two-thirds are clearly confirmed as celestial bodies (over 60 targets). The remaining detections are more difficult to distinguish due to their low brightness. We identified some spurious detections close to bright sources or at the edge of the detector. Contamination in the sample of 523 detections without optical counterparts in public surveys is here estimated to be of 4\%. They might be potential targets of interest, for instance, for studies on strong M-dwarf flares or on extreme stellar variability. The complete list of 523 detections will be accessible from the archive (see Section~\ref{sec:data_access} for details) and at CDS through the VizieR service.

\subsection{Identification of cool dwarfs}

M dwarfs are the most common stars in the Milky Way \citep{Kropua01,Chabrier03}. They are excellent tracers to study the structure, kinematics and evolution of the Galaxy \citep[e.g.,][]{Scalo86, Bochanski07, Ferguson17}. Also interesting is the fact that near one-third of M dwarfs are found in binary and multiple systems \citep[e.g.,][]{CC_FastCam}. Stellar companions located very close to the star are suitable for studying radii inflation \citep[e.g.,][]{Cruz2018,Cruz2022}.

We identified and characterised cool dwarfs to validate CAFOS photometry. The $r$ and $i$ filters were adopted, since they have the greatest number of detections among all filters (see Table~\ref{tab.cat-info}) and could provide information on the objects according to the colour ($r-i$), when both bands are available for the same object. Considering that only a small sample (over 3\,279 sources) have been observed with both filters, the selection was also based on Gaia photometry within 3\,arcsec, where we adopted the colours ($G-r$) and ($i-G_{\rm RP}$) to look for low temperature objects.

To select cool objects, we applied three different colour-cuts from the typical colours of M0V stars obtained in \citet{Cifuentes2020}. Among the sources with available CAFOS $r$ photometry, we kept those with available $G$ magnitudes with relative errors of less than 10\%. 
Then, we kept those sources with calculated $G-r$ colour of less than -0.17\,mag -- 3\,142 objects. Similarly, among the sources with detections in the $i$-band and with good $G_{\rm RP}$ magnitude (within 10\% error in Gaia magnitude), we selected the objects with ($i-G_{\rm RP}$)>0.36\,mag  -- 2\,941 objects. 
The last applied colour-cut considered all sources observed with both $r$ and $i$ filters. Among them, 584 objects presented ($r-i$) greater than 0.74\,mag. 
We then, selected a total of 5\,719 different sources which respected at least one of the above applied colour-cuts and that are therefore photometric candidates to M-type or later dwarfs. It is worth mentioning that 456 of the selected objects follow the three criteria. 

\begin{figure}
\centering
\includegraphics[width=0.95\columnwidth]{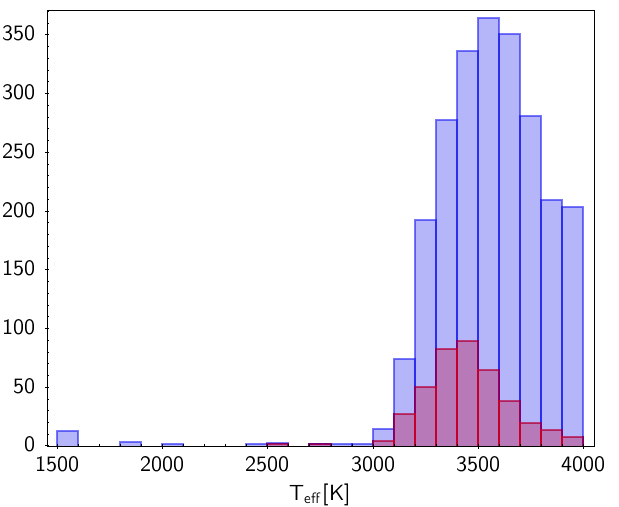}
\caption{Histogram of effective temperatures obtained from SED fitting with VOSA. The $T_{\rm eff}$ for the M-star sample (with 2\,322 objects) are presented in blue. The identified nearby cool dwarfs (within 500\,pc) are shown in red.}
\label{fig:Teff_Mstars}
\end{figure}

These M-star candidates underwent a second criterion, based on their effective temperatures. The $T_{\rm eff}$ were estimated from SED fitting obtained with VOSA. CAFOS $r$ and $i$ photometry were used to construct the SED -- also $g$ and $z$ band CAFOS magnitudes when available. We also searched for additional photometry available in the VO archives, covering the visible and infrared wavelengths, which are: SDSS DR12, Gaia eDR3, PanSTARSS DR2 \citep[][and references therein]{PS1DR2}, 2MASS, IRAS, AKARI/IRC \citep{AKARI}, and WISE. For the fit, we adopted the BT-Settl CIFIST models, with $T_{\rm eff}$ ranging from 1\,200 to 7\,000\,K, $\log g$ from 2.5 to 5.5\,dex, and solar metallicity ([Fe/H]=0.0). Any IR photometric data that seemed as an excess -- VOSA considers the photometric points which are above the best-fit model by over 3$\sigma$ to be an excess -- were excluded from the fit. 
From the list of candidates, 4\,822 objects presented very good SED fits -- based on the visual goodness of fit, Vgf$_{\rm b}$\footnote{For details, see VOSA help page.} parameter estimated by VOSA of less then $8$. 
We have then found 2\,322 objects with $T_{\rm eff}$ of less than 4\,000\,K, corresponding to M stars or later spectral types. The histogram of obtained temperatures is presented in blue in Figure \ref{fig:Teff_Mstars}. 
A cross match with the Washington Double Star (WDS) catalogue \citep{mason01} and the The Exoplanet Encyclopaedia\footnote{\url{www.exoplanet.eu}} indicates there are no counterparts in such catalogues. However, we found 58 M dwarfs in our sample with the {\it Gaia} {\tt RUWE} parameter above 1.4, which suggest a bad astrometric solution that may be ascribed to the presence of a close companion \citep{arenou18,lindegren18a}. One of them is an ultra cool dwarf of 1800\,K.

The remaining objects with good SED fitting (4\,822 - 2\,322) presented higher temperatures, consistent with K-type stars. We attributed this to: i) the initial selection was based on just one of the three applied colour-cuts; and ii) the ($i-G_{\rm RP}$) color is more effective on discriminating ultracool objects, with stellar types later than M6, rather than earlier types \citep[see Table A.2 from][for more details]{Cifuentes2020}.

\begin{figure}
\centering
\includegraphics[width=0.95\columnwidth]{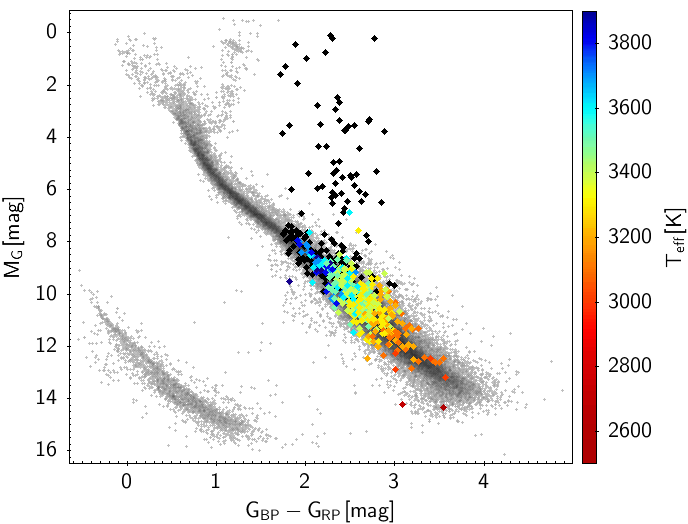}
\caption{Colour-magnitude diagram of identified M stars with good parallax measurements from Gaia EDR3. Nearby cool stars within 500\,pc are presented colour coded, according to their estimated temperatures. More distant objects, but with good parallaxes, are presented in black. A sample of Gaia EDR3 sources  with good parallax and proper motions are shown in grey.}
\label{fig:CMD}
\end{figure}

In order to find a subset of nearby M dwarfs, we selected those that presented good parallax measurements (with a parallax relative error of less than 10\%) and with calculated distances inferior to 500\,pc finding 395 objects. These stars are represented in a colour-magnitude diagram (CMD; Figure~\ref{fig:CMD}) by different colours, showing the obtained effective temperature. Those objects with good parallaxes, but with calculated distances beyond 500\,pc, are presented in black -- 601 stars. As reference, we included in the CMD a sample of Gaia EDR3 sources with relative errors lower than 10\% in parallax and proper motion, shown in grey. The obtained $T_{\rm eff}$ for these nearby objects are illustrated in Figure \ref{fig:Teff_Mstars} in red. 

Among our set of low-temperature objects, we identified 20 ultracool dwarfs -- that is, M7 type or later, with temperatures ranging from 1\,500 to 2\,700\,K, with associated types from M7 to L7. Three of them are included in the SIMBAD database, two of which are identified as an M8 dwarf and a brown dwarf candidate. Both of them lie within 500\,pc (91.4 and 102.8\,pc, respectively). The remaining candidates require spectroscopic confirmation. The list of ultracool dwarf candidates is shown in Table~\ref{tab.ucds} together with their measured CAFOS magnitudes in the $i$ band (or $r$ when $i$ is not available). The complete list of 2\,322 low mass stars candidates will be available in the archive (see Section~\ref{sec:data_access} for details), together with the sample of candidates at less than 500\,pc and those that satisfy the three colour criteria. They will also be accessible through the CDS VizieR service.

\begin{table}
\small
\centering
 \caption {Ultracool dwarf candidates identified in the catalogue.}
 \label{tab.ucds}
 \begin{tabular}{lcccl}
 \hline \hline
 \noalign{\smallskip}
Source NID & RA &   DEC &	$T_{\rm eff}$ & {i\_mag}   \\
 &   [hms]   &[dms]  & [K]  &  { [mag]}	\\

 \noalign{\smallskip}
 \hline
 \noalign{\smallskip}
 \noalign{\smallskip}
21884$^a$	&	01:28:50.58	&	+21:03:09.4	&	2\,700 &  20.77{$^c$} \\
7491$^b$	&	12:33:05.35	&	+44:55:34.9	&	2\,500 &   18.38 \\
11850	&	18:40:34.62	&	-00:02:26.5	&	1\,500 &   15.24  \\
16673	&	18:40:38.25	&	-00:01:08.2	&	1\,800 &    17.96\\
11928	&	18:40:40.21	&	+00:01:12.0	&	1\,500 &   16.32 \\
11991	&	18:40:45.46	&	+00:03:26.1	&	1\,500 &   17.53\\
12003	&	18:40:46.35	&	+00:04:53.7	&	1\,800 &   17.94 \\
12042	&	18:40:49.94	&	+00:05:06.8	&	1\,500 &   16.86 \\
16660	&	18:40:51.37	&	+00:03:36.6	&	2\,500 &   18.11\\
12079	&	18:40:53.19	&	+00:02:55.5	&	1\,500 &   16.84 \\
12101	&	18:40:54.58	&	+00:00:31.6	&	1\,500 &  13.98 \\
12120	&	18:40:56.34	&	+00:04:05.5	&	1\,800 &   17.21 \\
12129	&	18:40:56.90	&	+00:04:19.6	&	1\,500 &   17.07 \\
12147	&	18:40:58.56	&	+00:01:53.3	&	1\,500 &   15.63 \\
12161	&	18:41:00.01	&	+00:00:44.5	&	2\,400 &   14.49\\
12190	&	18:41:02.78	&	+00:01:20.4	&	1\,500 &   17.00\\
12193	&	18:41:02.85	&	-00:01:57.3	&	1\,500 &    16.73\\
12198	&	18:41:03.26	&	+00:03:01.5	&	1\,500 &   17.03\\
12232	&	18:41:06.58	&	-00:01:21.6	&	1\,500 &    16.75\\
16577	&	18:42:58.56	&	+00:27:36.3	&	2\,050 &    14.89$^c$\\

\noalign{\smallskip} 
 \hline
\end{tabular}
\begin{list}{}{}
\item[]{\scriptsize{ {\bf Note.} $^{(a)}$ Brown dwarf candidate \citep{Reyle18}. $^{(b)}$ Classified as M8V \citep{West11}. \hspace{1cm}$^{(c}$ Magnitudes in the $r$ filter.}}
\end{list}
\end{table}

\section{Data Access}\label{sec:data_access}

\begin{figure*}
\centering
\includegraphics[width=\textwidth]{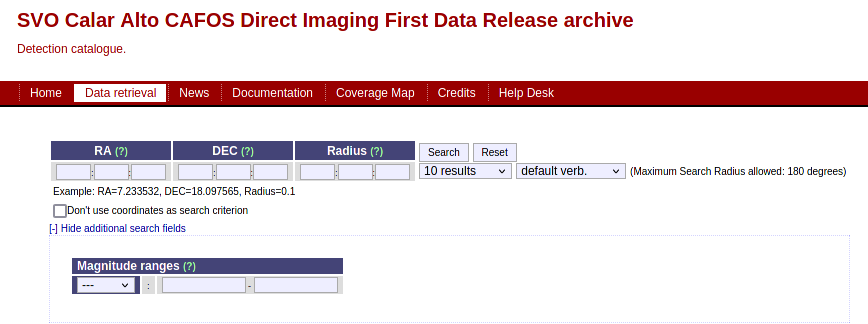}
\caption{Screenshot of the archive search interface that permits simple queries to the detection catalogue.}
\label{fig:svocat}
\end{figure*}

The astrometrically corrected CAFOS images are available from the CALAR ALTO Archive Portal\footnote{\url{https://caha.sdc.cab.inta-csic.es/calto/jsp/searchform.jsp}} as well as from the corresponding SIAP service\footnote{\url{http://caha.sdc.cab.inta-csic.es/calto/siap/caha_siap.jsp?}} while the catalogue can be accessed from an archive system\footnote{\url{http://svocats.cab.inta-csic.es/caha-cafos/}} or through a Virtual Observatory ConeSearch\footnote{e.g. \url{http://svocats.cab.inta-csic.es/caha-cafos-detection/cs.php?RA=356.979&DEC=32.346&SR=0.1&VERB=2}}.

This archive system provides a very simple search interface that allows queries by position or magnitude interval (Figure~\ref{fig:svocat}). It also implements a link to the reduced images in the CALAR Alto archive. 
After launching the query, the result is presented in HTML format with all the detections that satisfy the search criteria. The output can also be downloaded as a VOTable or CSV file. The results can also be easily transferred to other VO applications like TOPCAT \citep{Taylor05} through the SAMP protocol.
Detailed information on the output fields can be obtained placing the mouse over the
question mark ("?") located near the name of the column.

\section{Conclusions and future work}\label{sec:conclusions}

In this work, we presented the first data release of the Calar Alto CAFOS direct imaging data, which includes on one hand, 42\,173 ready-to-use processed and astrometrically calibrated images from March 2008 to July 2019. On the other hand, it provides the photometric catalogue associated to 6\,132 images observed in Sloan's $griz$ filters. It contains 139\,337 point-like detections corresponding to 21\,985 different sources 
taken between September 2012 and June 2019. Exposure times in the catalogue range from 0.5 to 600\,s.

We developed the CAFOS Photometry Calibrator ({\tt CFC}) with the goal of automatically extracting the sources from the images, carrying out the selection of valid non-saturated sources and performing the flux calibration with the SDSS DR12 or APASS DR9 surveys as reference. 

The final catalogue of point sources has a mean internal astrometric accuracy of 0.05\,arcsec.
The mean internal photometric accuracy is 0.04\,mag and the averaged standard deviation of magnitudes of sources detected more than five times in each filter is of 0.08\,mag. No dependence on colour, binning mode or pixel position was found. Additionally, we conducted an exercise to assess the presence of real variable objects in the catalogue and obtained a sample of 209 targets suitable for a variability analysis.

To demonstrate the science capabilities of the photometric catalogue, we explore three different science cases. We first searched for SSOs that serendipitously cross the FoV of the images. We recovered 446 positions in 341 images of 20 SSOs, seven of which are unknown. These results have been shared with the Minor Planet Center. Secondly, we inspected a sample of 516 sources without counterparts in the {\it Gaia} EDR3, SDSS DR12 or Pan-STARRS DR1 optical surveys. One of them is a known gamma-ray burst, another 22 have counterparts in the ZTF archive and near 40\% (203 objects) have NEOWISE photometry, meaning there are very red objects in the sample. For six targets, we complemented our optical magnitudes with infrared photometry and provided effective temperatures between 2\,400 and 4\,100\,K from their SED analysis using VOSA. These are estimated K-M dwarfs that might happen to flare during the observations.
Furthermore, near 40\% are related to moving SSOs recovered previously with the {\tt ssos} pipeline. Around 60 targets lack of further information for a proper characterization and are proposed as potential targets of interest for studies on M dwarf flares or stellar variability, for instance. 
At last, we looked for cool dwarfs in the catalogue from the application of three colour cuts in the optical wavelength range using photometry in the $r$ and $i$ bands together with {\it Gaia}'s. We identified 2\,322 sources with effective temperatures ranging between 1\,500 and 3\,900\,K obtained with VOSA, 326 of which satisfy the three colour criteria simultaneously. 58 M dwarfs have a {\it Gaia}'s {\tt RUWE} parameter over 1.4, which is related to a bad astrometric solution possibly ascribed to the presence of a close companion. Out of the 2\,322 low mass stars candidates, 20 are ultra cool dwarfs that present temperatures under 2\,700\,K. Additionally, we defined a subset of 395 nearby cool dwarfs with distances closer than 500\,pc.

For new releases of the catalogue we aim at including rejected images in this work and images observed in filters different than the Sloan's $griz$ bands. Also new public broad-band images that will be delivered over the life of the instrument would be included. In order to improve the photometric quality of the catalogue, we plan to test the photometric calibration by using the ATLAS All-Sky Stellar Reference Catalog \citep{Tonry2018} as well as Gaia spectra from which we could derive photometry without requiring further corrections.

Given the heterogeneity of the observing conditions and purposes of the images, and despite the homogeneus treatment applied in this work, the complete absence of errors and problems in the catalogue cannot be guaranteed. Users are therefore encouraged in these cases to download and check the images to assess the reliability of a given catalogue measurement. The {\it News} section of the catalogue website will contain a list of Frequently Asked Questions as well as a description of caveats that may arise with the scientific exploitation of the catalogue.

\section*{Acknowledgments}

This research is part of the proyect I+D+i with reference PID2020-112949GB-I00, supported by the MCIN/ AEI/10.13039/501100011033 and MDM-2017-0737 at Centro de Astrobiolog\'{i}a (CSIC-INTA), Unidad de Excelencia Mar\'{i}a de Maeztu, and the ESCAPE project supported by the European Commission Framework Programme Horizon 2020 Research and Innovation action under grant agreement n. 824064.
NC acknowledges financial support from the Spanish Programa Estatal de I+D+i Orientada a los Retos de la Sociedad under grant RTI2018-096188-B-I00, which is partly funded by the European Regional Development Fund (ERDF).
PC acknowledges financial support from the Government of Comunidad Autónoma de Madrid (Spain), via postdoctoral grant ‘Atracción de Talento Investigador’ 2019-T2/TIC-14760.
MCC thanks Max Mahlke for his valuable support in the {\tt ssos} usage.
This research has made use of the Spanish Virtual Observatory (http://svo.cab.inta-csic.es) supported from Ministerio de Ciencia e Innovación through grant PID2020-112949GB-I00.
This publication makes use of VOSA, developed under the Spanish Virtual Observatory project supported by the Spanish MINECO through grant AyA2017-84089. VOSA has been partially updated by using funding from the European Union's Horizon 2020 Research and Innovation Programme, under Grant Agreement nº 776403 (EXOPLANETS-A).
This research has made use of the SVO Filter Profile Service (http://svo2.cab.inta-csic.es/theory/fps/) supported from the Spanish MINECO through grant AYA2017-84089.
This research has made use of the cross-match service, the SIMBAD database \citep{Wenger00}, VizieR catalogue access tool \citep{Och00}, "Aladin sky atlas" \citep{Bonn00, BF14} provided by CDS, Strasbourg, France.
This research has also made use of the TOPCAT \citep{Taylor05} and STILTS \citep{Taylor06}.
This research has made use of IMCCE's SkyBoT VO tool \citep{Ber06}, {\tt ssos} \citep{Mahlke2019}, {\sc SExtractor} \citep{1996A&AS..117..393B}, {\sc PSFEx} \citep{Bertin11}, {\tt SCAMP} \citep{scamp}.
This research has made use of the NASA/IPAC Infrared Science Archive, which is funded by the National Aeronautics and Space Administration and operated by the California Institute of Technology. 
This research has made use of the International Variable Star Index (VSX) database, operated at AAVSO, Cambridge, Massachusetts, USA.
This research was made possible through the use of the AAVSO Photometric All-Sky Survey (APASS), funded by the Robert Martin Ayers Sciences Fund and NSF AST-1412587.
This work has made use of data from the European Space Agency (ESA) mission {\it Gaia} (\url{https://www.cosmos.esa.int/gaia}), processed by the {\it Gaia} Data Processing and Analysis Consortium (DPAC, \url{https://www.cosmos.esa.int/web/gaia/dpac/consortium}). Funding for the DPAC has been provided by national institutions, in particular the institutions participating in the {\it Gaia} Multilateral Agreement.
This work has been possible thanks to the extensive use of IPython and Jupyter notebooks \citep{PER-GRA:2007}, as well as the Python packages {\sc astropy}\footnote{\url{http://www.astropy.org}}, a community-developed core Python package for Astronomy \citep{astropy13, astropy18}, {\sc astroquery} \citep{astroquery}, {\sc numpy} \citep{numpy20}, {\sc scipy} \citep{scipy20}, and {\sc matplotlib} \citep{matplotlib07}.

\section{Data availability}

The data underlying this article are available in the SVO Calar Alto CAFOS Direct Imaging First Data Release archive at http://svocats.cab.inta-csic.es/caha-cafos/

\bibliographystyle{mnras}
\bibliography{biblio.bib}

\clearpage
\onecolumn
\appendix

\newpage

\section{Configuration files}\label{app.files}

\centering
\begin{longtable}{lr}
\caption{Configuration parameters for the first iteration of {\sc SExtractor}.}\\
\label{tab.def_sex}\\
\hline
\noalign{\smallskip}
\endfirsthead
\caption{Configuration parameters for the first iteration of  {\sc SExtractor} (continued).}\\
\hline
\noalign{\smallskip}
\endhead
\multicolumn{2}{c}{Catalog}	\\
\noalign{\smallskip}
\hline 
 \noalign{\smallskip}

CATALOG\_NAME$^a$ &   CFC\_configuration/sextractor\_result\_files/test\_psf.cat     \\  

CATALOG\_TYPE$^b$ & FITS\_LDAC \\
PARAMETERS\_NAME & CFC\_configuration/sextractor\_configuration\_files/prepsfex.param \\ 

\noalign{\smallskip}
\hline
\noalign{\smallskip}
\multicolumn{2}{c}{Extraction}  \\
\noalign{\smallskip}
\hline
\noalign{\smallskip}

DETECT\_TYPE  &   CCD         \\
DETECT\_MINAREA & 12      \\
DETECT\_MAXAREA & 0\\
THRESH\_TYPE  &  RELATIVE \\%
DETECT\_THRESH  & 1.5    \\
ANALYSIS\_THRESH & 1.5         \\

FILTER       &   Y \\
FILTER\_NAME  &   CFC\_configuration/sextractor\_configuration\_files/gauss\_2.5\_5x5.conv \\

DEBLEND\_NTHRESH &32       \\
DEBLEND\_MINCONT &0.005  \\

CLEAN        &   Y       \\
CLEAN\_PARAM   &  1.0    \\

MASK\_TYPE     &  CORRECT   \\

\noalign{\smallskip}
\hline
\noalign{\smallskip}
\multicolumn{2}{c}{Photometry}  \\
\noalign{\smallskip}
\hline
\noalign{\smallskip}

PHOT\_APERTURES & 20,25,30    \\
PHOT\_AUTOPARAMS & 2.5, 3.5   \\
PHOT\_PETROPARAMS & 1.0,3.5\\%

PHOT\_AUTOAPERS & 0.0,0.0\\%
PHOT\_FLUXFRAC & \\
SATUR\_LEVEL   &  50000.0    \\
SATUR\_KEY & DUMMY\\%
MAG\_ZEROPOINT &  0.0\\
MAG\_GAMMA    &   4.0 \\
GAIN        &    0.0 \\
GAIN\_KEY & GAIN\\%
PIXEL\_SCALE   &  0\\

\noalign{\smallskip}
\hline
\noalign{\smallskip}
\multicolumn{2}{c}{Star/Galaxy Separation}  \\
\noalign{\smallskip}
\hline
\noalign{\smallskip}
SEEING\_FWHM $^c$ &   1.2         \\
STARNNW\_NAME  &  CFC\_configuration/sextractor\_configuration\_files/default.nnw  \\
 
\noalign{\smallskip}
\hline
\noalign{\smallskip}
\multicolumn{2}{c}{Background}  \\
\noalign{\smallskip}
\hline
\noalign{\smallskip}
  
BACK\_SIZE &      64  \\
BACK\_FILTERSIZE &5       \\
 
BACKPHOTO\_TYPE  &GLOBAL     \\
 
\noalign{\smallskip}
\hline
\noalign{\smallskip}
\multicolumn{2}{c}{Check Image} \\
\noalign{\smallskip}
\hline
\noalign{\smallskip}
  
CHECKIMAGE\_TYPE$^d$ & NONE     \\
CHECKIMAGE\_NAME$^e$ & check.fits,aper.fits   \\
 
\noalign{\smallskip}
\hline
\noalign{\smallskip}
\multicolumn{2}{c}{Memory}   \\
\noalign{\smallskip}
\hline
\noalign{\smallskip} 
MEMORY\_OBJSTACK &30000 \\
MEMORY\_PIXSTACK &300000   \\
MEMORY\_BUFSIZE & 1800        \\
 
\noalign{\smallskip}
\hline
\noalign{\smallskip}
\multicolumn{2}{c}{Miscellaneous}   \\
\noalign{\smallskip}
\hline
\noalign{\smallskip}

WEIGHT\_TYPE & NONE \\%
VERBOSE\_TYPE   & NORMAL       \\
HEADER\_SUFFIX$^f$ & .head \\
INTERP\_MAXXLAG$^f$ & 16\\
INTERP\_MAXYLAG$^f$ & 16\\
INTERP\_TYPE $^f$ &NONE\\
PSF\_NAME $^f$ &CFC\_configuration/sextractor\_result\_files/test\_psf.psf\\
PATTERN\_TYPE$^f$ & RINGS-HARMONIC\\
SOM\_NAME $^f$ &default.som\\

\noalign{\smallskip}
\hline
\noalign{\smallskip}

\begin{minipage}{0.5\textwidth}
\vspace{0pt}
{\footnotesize{\textbf{Notes}.
$^{(a)}$ {Catalogue name was renamed to "test.cat" for the second iteration of {\sc SEXtractor}.}
$^{(b)}$ {Catalogue type was changed to "ASCII\_HEAD" for the second iteration of {\sc SEXtractor}.}
$^{(c)}$ {Stellar seeing was modified to 2.5 for the second iteration of {\sc SEXtractor}.}
$^{(d)}$ {Checkimage type was modified to "BACKGROUND" for the second iteration of {\sc SEXtractor}.}
$^{(e)}$ {Checkimage name was modified to "CFC\_configuration/sextractor\_result\_files/bkg.fits"}. \\
$^{(f)}$ {Only included in the configuration file of the second iteration of {\sc SEXtractor}.}
}}
\end{minipage}

\end{longtable}

\newpage
\centering
\begin{longtable}{lr}
\caption{Configuration paramerers for {\sc PSFEx}.}\\
\label{tab.def_psfex}\\
\hline
\noalign{\smallskip}
\endfirsthead
\caption{Configuration paramerers for {\sc PSFEx} (continued).}\\
\hline
\noalign{\smallskip}
\endhead
\multicolumn{2}{c}{PSF model}	\\
 \noalign{\smallskip}
 \hline
 \noalign{\smallskip}

BASIS\_TYPE   &   PIXEL\_AUTO \\
BASIS\_NUMBER  &  20            \\
BASIS\_NAME    &  basis.fits     \\
BASIS\_SCALE   &  1.0          \\
NEWBASIS\_TYPE &  NONE        \\
NEWBASIS\_NUMBER& 8        \\
PSF\_SAMPLING  &  0.0         \\
PSF\_PIXELSIZE &  1.0        \\
PSF\_ACCURACY  &  0.01       \\
PSF\_SIZE     &   31,31       \\
PSF\_RECENTER  &  Y           \\
MEF\_TYPE     &   INDEPENDENT   \\
 
 \noalign{\smallskip}
\hline
\noalign{\smallskip}

\multicolumn{2}{c}{Point source measurements}	\\ 
\noalign{\smallskip}
\hline
\noalign{\smallskip}
 
CENTER\_KEYS    & X\_IMAGE,Y\_IMAGE\\
PHOTFLUX\_KEY   & FLUX\_APER(1)   \\
PHOTFLUXERR\_KEY& FLUXERR\_APER(1)\\
 
  \noalign{\smallskip}
\hline
\noalign{\smallskip}

\multicolumn{2}{c}{ PSF variability}	\\ 
\noalign{\smallskip}
\hline
\noalign{\smallskip}

PSFVAR\_KEYS   &  X\_IMAGE,Y\_IMAGE \\
PSFVAR\_GROUPS &  1,1           \\
PSFVAR\_DEGREES&  3               \\
PSFVAR\_NSNAP  &  9          \\
HIDDENMEF\_TYPE  &COMMON        \\
STABILITY\_TYPE  &EXPOSURE        \\
 
 \noalign{\smallskip}
\hline
\noalign{\smallskip}

\multicolumn{2}{c}{Sample selection}	\\ 
\noalign{\smallskip}
\hline
\noalign{\smallskip}

SAMPLE\_AUTOSELECT & Y        \\
SAMPLEVAR\_TYPE   &  SEEING  \\
SAMPLE\_FWHMRANGE  & 1.0,15.0 \\
SAMPLE\_VARIABILITY &0.05    \\
SAMPLE\_MINSN     &  20       \\
SAMPLE\_MAXELLIP  &  0.3      \\
SAMPLE\_FLAGMASK  &  0x00fe   \\
SAMPLE\_WFLAGMASK  & 0x00ff   \\
SAMPLE\_IMAFLAGMASK &0x0      \\
BADPIXEL\_FILTER   & N         \\
BADPIXEL\_NMAX   &   0         \\
 
 \noalign{\smallskip}
\hline
\noalign{\smallskip}

\multicolumn{2}{c}{PSF homogeneisation kernel}	\\ 
\noalign{\smallskip}
\hline
\noalign{\smallskip}

HOMOBASIS\_TYPE   &  NONE     \\
HOMOBASIS\_NUMBER &  10       \\
HOMOBASIS\_SCALE &   1.0       \\
HOMOPSF\_PARAMS  &   2.0, 3.0  \\
HOMOKERNEL\_DIR   &            \\
HOMOKERNEL\_SUFFIX & .homo.fits  \\

 \noalign{\smallskip}
\hline
\noalign{\smallskip}

\multicolumn{2}{c}{Output catalogs}	\\ 
\noalign{\smallskip}
\hline
\noalign{\smallskip}

OUTCAT\_TYPE   &     NONE       \\
OUTCAT\_NAME   &     CFC\_configuration/sextractor\_result\_files/psfex\_out.cat \\

 \noalign{\smallskip}
\hline
\noalign{\smallskip}

\multicolumn{2}{c}{Check-plots}	\\ 
\noalign{\smallskip}
\hline
\noalign{\smallskip}

CHECKPLOT\_DEV   &    NULL     \\
CHECKPLOT\_RES    &   0        \\
CHECKPLOT\_ANTIALIAS &Y        \\
CHECKPLOT\_TYPE &     NONE \\
CHECKPLOT\_NAME  &    NONE \\
 
 \noalign{\smallskip}
\hline
\noalign{\smallskip}

\multicolumn{2}{c}{Check-Images}	\\ 
\noalign{\smallskip}
\hline
\noalign{\smallskip}

CHECKIMAGE\_TYPE& NONE \\
CHECKIMAGE\_NAME& chi.fits,proto.fits,samp.fits,resi.fits,snap.fits\\%
CHECKIMAGE\_CUBE N  &       \\
 
 \noalign{\smallskip}
\hline
\noalign{\smallskip}

\multicolumn{2}{c}{Miscellaneous}	\\ 
\noalign{\smallskip}
\hline
\noalign{\smallskip}

PSF\_DIR     &                \\
PSF\_SUFFIX  &    .psf         \\
VERBOSE\_TYPE  &  NORMAL       \\
WRITE\_XML   &    N            \\
XML\_NAME    &    psfex.xml     \\
XSL\_URL     &    file:///usr/share/psfex/psfex.xsl\\%
NTHREADS      &  1           \\

 \noalign{\smallskip}
\hline
\end{longtable}

\newpage
\centering
\begin{table}
\caption{Parameters obtained from {\sc SExtractor} and {\sc PSFEx}.}
\label{tab.def_params}
\begin{tabular}{lll}
\hline\hline
\noalign{\smallskip}

NUMBER & Running object number     \\
ALPHA\_J2000	&	Right ascension of barycenter (J2000)           &          [deg]	\\
DELTA\_J2000	&	Declination of barycenter (J2000)                &         [deg]	\\

X\_IMAGE	&	Object position along x                &                   [pixel]	\\
Y\_IMAGE	&	Object position along y                 &                  [pixel]	\\
A\_IMAGE	&	Profile RMS along major axis             &                 [pixel]	\\
B\_IMAGE	&	Profile RMS along minor axis              &                [pixel]	\\
THETA\_IMAGE	&	Position angle (CCW/x)                     &               [deg]	\\

XWIN\_IMAGE	&	Windowed position estimate along x                  &      [pixel]	\\
YWIN\_IMAGE	&	Windowed position estimate along y                   &     [pixel]	\\
XMIN\_IMAGE     &          Minimum x-coordinate among detected pixels          &      [pixel]	\\
YMIN\_IMAGE      &         Minimum y-coordinate among detected pixels           &     [pixel]	\\
XMAX\_IMAGE       &        Maximum x-coordinate among detected pixels            &    [pixel]	\\
YMAX\_IMAGE        &       Maximum y-coordinate among detected pixels             &   [pixel]	\\

X\_WORLD	&	Barycenter position along world x axis            &        [deg]	\\
Y\_WORLD	&	Barycenter position along world y axis             &       [deg]	\\
ERRX2\_WORLD	&	Variance of position along X-WORLD (alpha)          &      [deg**2]	\\
ERRY2\_WORLD	&	Variance of position along Y-WORLD (delta)           &     [deg**2]	\\

FLAGS	&   Extraction flags	&   	\\
FLAGS\_WEIGHT	&   Weighted extraction flags	&   	\\

FWHM\_IMAGE	&	FWHM assuming a gaussian core               &              [pixel]	\\
FWHM\_WORLD	&	FWHM assuming a gaussian core                     &        [deg]	\\

FLUX\_MAX	&	Peak flux above background                   &             [count] \\
FLUX\_RADIUS	&	Fraction-of-light radii                       &            [pixel]	\\
FLUX\_PSF	&	Flux from PSF-fitting                              &       [count]	\\
FLUXERR\_PSF	&	RMS flux error for PSF-fitting                      &      [count]	\\
MAG\_PSF	&	Magnitude from PSF-fitting                           &     [mag]	\\
MAGERR\_PSF	&	RMS magnitude error from PSF-fitting                  &    [mag]	\\
FLUX\_AUTO        &        Flux within a Kron-like elliptical aperture        &       [count]	\\
FLUXERR\_AUTO      &       RMS error for AUTO flux                             &      [count]	\\
MAG\_AUTO           &      Kron-like elliptical aperture magnitude              &     [mag]	\\
MAGERR\_AUTO         &     RMS error for AUTO magnitude                          &    [mag]	\\

ELONGATION	&	A\_IMAGE/B\_IMAGE where A\_IMAGE is the profile RMS along the major &   \\
        &   axis and B\_IMAGE is the profile RMS along the minor axis&	\\
ELLIPTICITY	&	1 - B\_IMAGE/A\_IMAGE (where A\_IMAGE is the profile RMS along the major    \\
    & axis and B\_IMAGE is the profile RMS along the minor axis	& 	\\
SNR\_WIN	&	Signal-to-noise ratio in a Gaussian window	&	\\
SPREAD\_MODEL	&	Spread parameter from model-fitting	&	\\

\noalign{\smallskip}
\hline
\end{tabular}
\end{table}

\newpage

\section{Catalogue description}\label{app.cats}

\centering
\begin{table}
\caption{Description of the parameters contained in the source catalogue.}
\label{tab.catalogue_description}
\begin{tabular}{p{2.5cm}p{1cm}p{13cm}}
\hline \hline
	\noalign{\smallskip}
Parameter    &   Units   &   Description \\
	\noalign{\smallskip}

	\hline
SourceNID	&	-	&	Numerical ID of sources detected more than once within 0.5 arcsec	\\
RAJ2000	&	deg	&	Celestial Right Ascension (J2000)	\\
DEJ2000	&	deg	&	Celestial Declination (J2000)	\\
e\_RAJ2000	&	arcsec	&	Right ascension error	\\
e\_DEJ2000	&	arcsec	&	Declination error	\\
Detection\_ID	&	-	&	Detection identifier	\\
Image\_identifier	&	-	&	Image identifier	\\
Image\_url	&	-	&	URL to access the reduced image	\\
gmag	&	mag	&	PSF calibrated magnitude in the g band	\\
e\_gmag	&	mag	&	PSF calibrated magnitude error in the g band	\\
rmag	&	mag	&	PSF calibrated magnitude in the r band	\\
e\_rmag	&	mag	&	PSF calibrated magnitude error in the r band	\\
imag	&	mag	&	PSF calibrated magnitude in the i band	\\
e\_imag	&	mag	&	PSF calibrated magnitude error in the i band	\\
zmag	&	mag	&	PSF calibrated magnitude in the z band	\\
e\_zmag	&	mag	&	PSF calibrated magnitude error in the z band	\\
Flag\_phot	&	-	&	Calibration flag. "A" stands for calibrated magnitudes within the interval of magnitudes used for the photometric calibration, "B" for calibrated magnitudes fainter than the faintest magnitudes, and "C" for calibrated magnitudes brighter magnitudes than the brightest magnitude used for the photometric calibration.		\\	
Filter	&	-	&	Filter of observation	\\
Refcat	&	-	&	Reference catalogue used for the photometric calibration	\\
Elongation	&	-	&	Elongation of the detection	\\
Ellipticity	&	-	&	Ellipticity of the detection	\\
FWHM	&	arcsec	&	Full width half maximum of the detection	\\
SNR\_WIN	&	-	&	Signal to noise ratio	\\
SPREAD\_MODEL	&	-	&	Spread parameter from model-fitting. It takes values close to zero for point sources, positive for extended sources (galaxies), and negative for detections smaller than the PSF, such as cosmic ray hits.	\\
MJD	&	d	&	Modified Julian Date	\\
DATE-OBS	&	iso-8601	&	Observing epoch	\\
SourceSize	&	-	&	Number of detections for each source in GroupID	\\

	\noalign{\smallskip}

\hline
\end{tabular}
\end{table}



\label{lastpage}
\end{document}